\documentclass[final,oneeqnum,onefignum,onetabnum]{siamltex}

\usepackage{graphicx,amsmath,amssymb}

\usepackage{cite}
\usepackage{wasysym}

\newcommand{\tsfrac}[2]{{\textstyle \frac{#1}{#2}}}

\newcommand{\lambdastar}{\lambda^\dagger}
\newcommand{\sineparam}{T}

\graphicspath{{Figures/}}

\title{Topological entropy of braids on the torus\thanks{Department of
    Mathematics, Imperial College London, London SW7 2AZ, United Kingdom}}

\author{Matthew D. Finn\thanks{{\tt matthew.finn@imperial.ac.uk}} \and Jean-Luc
    Thiffeault\thanks{{\tt jeanluc@imperial.ac.uk}}}

\begin{document}

\maketitle

\begin{abstract}
A fast method is presented for computing the topological entropy of braids on
the torus. This work is motivated by the need to analyze large braids when
studying two-dimensional flows via the braiding of a large number of particle
trajectories. Our approach is a generalization of Moussafir's technique for
braids on the sphere. Previous methods for computing topological entropies
include the Bestvina--Handel train-track algorithm and matrix representations
of the braid group.  However, the Bestvina--Handel algorithm quickly becomes
computationally intractable for large braid words, and matrix methods give only
lower bounds, which are often poor for large braids. Our method is
computationally fast and appears to give exponential convergence towards the
exact entropy. As an illustration we apply our approach to the braiding of 
both periodic and aperiodic trajectories in the sine flow. The efficiency of 
the method allows us to explore how much extra information about flow entropy 
is encoded in the braid as the number of trajectories becomes large.
\end{abstract}



\pagestyle{myheadings}
\thispagestyle{plain}
\markboth{M. D. FINN AND J.-L. THIFFEAULT}
	 {TOPOLOGICAL ENTROPY OF BRAIDS ON THE TORUS}

\section{Introduction}

Investigation of two-dimensional fluid mixing by topological techniques is
rapidly gaining popularity~\cite{Aref2002}. Topological perspectives on mixing
either involve studying braiding motion of the stirring apparatus
itself~\cite{Boyland2000}, or the diagnosis of mixing by analyzing braiding of
orbits of the flow
\cite{Gambaudo1999,Thiffeault2005,Gouillart2006,Finn2005}. The quantity that
is usually of interest is the topological entropy of the
braid~\cite{Boyland1994}, which serves as a lower bound for the topological
entropy of the flow. The topological entropy of the flow is related to the
exponential growth rate of material lines~\cite{Newhouse1993}, which has long
been a favorite measure of mixing quality, though it is by no means the only
one~\cite{Finn2004a}. In many cases the braid entropy is quite a sharp bound
on the flow entropy~\cite{Finn2003a,Finn2005,Gouillart2006}, which is one
reason why analyzing braids is useful. Another reason is that experimental
particle trajectory data can be found easily by particle image velocimetry,
but it is usually very difficult to measure entropies directly from material
stretching or by computing Lyapunov exponents.

There are many techniques for calculating braid topological entropies, or
lower bounds on them, including train-tracks~\cite{Bestvina1995,HallTrain},
the Burau representation of the braid group~\cite{Kolev1989}, and others
\cite{Lefranc2005,Moussafir2006}.  Where braiding of periodic (or aperiodic)
orbits is used to analyze a flow one needs to interpret braids that have a
both a large number of strands and a large number of generators. In this
scenario exact methods based on train-tracks quickly become prohibitively
expensive computationally, and methods based on the Burau matrix
representation of the braid group usually give very poor lower
bounds. Recently, however, Moussafir described a fast
method for calculating the entropy of a braid to an arbitrary precision
\cite{Moussafir2006}. This method is based on a Dynnikov coordinate
representation of a lamination~\cite{Dynnikov2002}. In this paper we show how
to extend Moussafir's technique to braids on the torus.  We are motivated by
the fact that many interesting dynamical systems are defined in periodic
(cylindrical or annular) or biperiodic (toroidal) spatial domains.  For
instance, the alternating sine flow~\cite{Finn2005} can be analyzed from a
topological perspective using this approach.

\begin{figure}
\begin{center}
\includegraphics{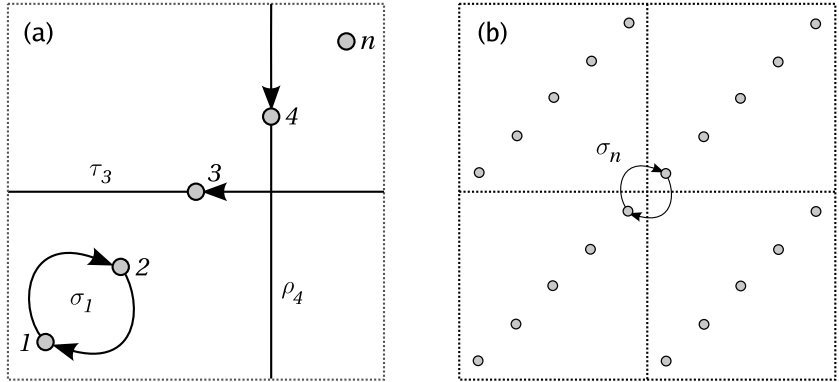}
\end{center}
\caption{The spatially periodic domain with $n$ punctures.  (a) Any motion of
the punctures (up to homotopy) can be written as a braid word consisting of a
string of $\sigma_i$, $\rho_i$ and $\tau_i$ motions and their inverses. (b) It
is convenient to define the additional operation~$\sigma_n$, the clockwise
exchange of the last and first punctures.}
\label{fig:bsd}
\end{figure}

Figure~\ref{fig:bsd} shows the setting for the dynamical system under study.
We consider a flow on a torus, so that the domain is periodic in both
directions.  We identify~$n$ distinguished points that we call punctures.
We then consider motions that move the punctures in such a way that they
always return to their initial configuration, possibly having been permuted
amongst themselves.  The three types of motions that we consider are
illustrated in Figure~\ref{fig:bsd}(a); they are
\begin{enumerate}
\item $\sigma_i$, the clockwise interchange of the $i$th and $(i+1)$th
  puncture;
\item $\rho_i$, the $i$th puncture making a full tour around the vertical
  periodic direction;
\item $\tau_i$, the $i$th puncture making a full tour around the horizontal
  periodic direction.
\end{enumerate}

The inverse of any of these motions is obtained by reversing its
direction. The elementary motions~$\{\sigma_i,\rho_i,\tau_i\}$ are generators
of the braid group on~$n$ strands on the torus~\cite{Birman1969}. Recognizing
the periodicity of the domain, we also define an additional operation
$\sigma_n$ to be the clockwise interchange of the~$n$th puncture with the
first puncture. To be precise, we mean here the first puncture in the `copy'
of the domain above and to the right of the $n$th puncture, as pictured in
Figure~\ref{fig:bsd}(b), so that both periodic boundaries are crossed in
performing $\sigma_n$.  Defining $\sigma_n$ in this way keeps both periodic
directions on an equal footing, and is also convenient in what follows as it
is related to a translational symmetry for the punctures.

A sequence of generators, such as $\rho_3^{-1} 
\sigma_2 \rho_6 \sigma_6$, is called a braid word, and we use the convention 
that the elementary motions in a braid word occur from left to right, so 
that~$\rho_3^{-1}$ occurs first in our example. 
By \emph{planar braid} we mean a braid word that can be
written using only generators from the set~$\{\sigma_1,\ldots,\sigma_{n-1}\}$,
and their inverses.  A planar braid is equivalent to a braid on the plane with
$n$ punctures.  In other words, a planar braid does not take advantage of the
periodic directions.  By \emph{cylinder braid} (or annular braid) we mean a 
braid word that can be written using only generators from either the
set~$\{\sigma_1,\ldots,\sigma_{n-1},\rho_1,\ldots,\rho_n\}$
or~$\{\sigma_1,\ldots,\sigma_{n-1},\tau_1,\ldots,\tau_n\}$, and their
inverses.  In other words, a cylinder braid takes advantage of one periodic
direction, but not the other.  Finally, a \emph{torus braid} is a braid word 
that is neither a planar braid nor a cylinder braid.  In this paper we will 
derive a general method for torus braids, which includes planar and cylinder 
braids as special cases.

The description of our method here is intended to be accessible to dynamicists
and requires no specialist understanding of braid groups. The paper is divided
as follows. In the following section we describe how laminations (equivalence
classes of simple closed curves) can be encoded by triangulation of the flow
domain. Section~\ref{sec:deformation}, the heart of the paper, gives the
details of how this encoding evolves under fundamental braiding actions or
motions.
Several examples are given in Section~\ref{sec:examples} to
illustrate and verify the method. In Section~\ref{sec:entropy} we show how to
compute the topological entropy from evolution of laminations, and demonstrate
the rapid convergence. We also look at braiding in the sine flow as an example
application. We summarize our work in Section~\ref{sec:discussion} and discuss
some features and possibilities for further study.

\section{Encoding of laminations by triangulation}
\label{sec:encoding}

\begin{figure}
\begin{center}
\includegraphics{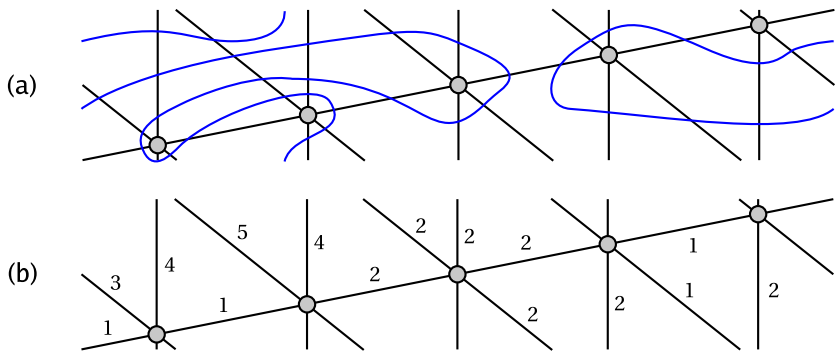}
\end{center}
\caption{(a) A representative curve of a lamination on a toroidal domain with
five punctures.  The top and bottom edges, and the left and right edges, are
identified so that the picture shows exactly ten triangles. (b) The number of
crossings with each edge of the triangulation, from which the lamination can
be reconstructed.}
\label{fig:puzzle}
\end{figure}

We wish to calculate a lower bound on how rapidly material lines are stretched
in a continuous-time flow based on the motion of a finite number of punctures,
or a finite set of periodic orbits. Our approach, inspired by the method of
Moussafir~\cite{Moussafir2006}, is to study the stretching and folding of
laminations. A lamination is an equivalence class (under homotopy) of simple
closed curves that are not homotopic to any part of the boundary (treating the
punctures as boundaries), and cannot be continuously shrunk to a point. For
example, a loop that encloses at least two punctures belongs to a lamination.
We usually represent a lamination by drawing one loop in the equivalence
class.  An example of a lamination on our doubly-periodic toroidal domain is
shown in Figure~\ref{fig:puzzle}(a).

\begin{figure}
\begin{center}
\includegraphics{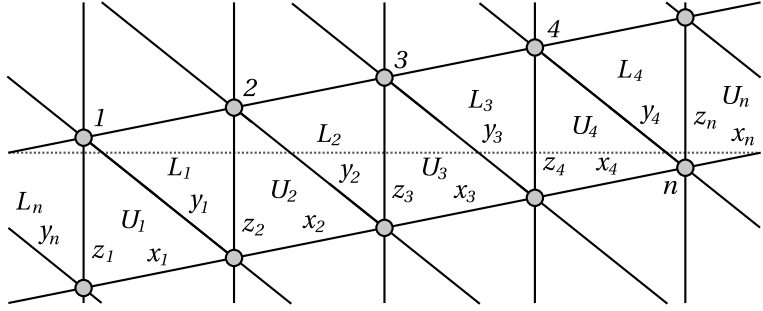}
\end{center}
\caption{Notation for our chosen triangulation of the doubly periodic domain
with $n$ punctures. The domain is divided into $2n$ triangles $U_i$ and $L_i$.
Two copies of the domain are shown one above the other for clarity. The $3n$
numbers $x_i$, $y_i$ and $z_i$ are used to count how many crossings a
lamination makes with each edge of the triangulation. The lamination is
assumed to be pulled tight, which means that when the curve enters a triangle
it must leave through a different edge to the one it entered by. Note that
this triangulation is not unique, but is chosen for convenience in the
calculations that follow.}
\label{fig:triangulation}
\end{figure}

In order to calculate how a given lamination is transformed under the action
of the braid, it is necessary to have a way of encoding the lamination, and a
method for evolving this encoding under the action of the braid.  An elegant
way of encoding the lamination is by triangulating the entire domain and
counting the number of crossings the lamination makes with the edges of the
triangulation. The number of crossings for our example lamination is shown in
Figure~\ref{fig:puzzle}(b). Note that there are infinitely many different ways
the domain can be triangulated, so in this paper we choose a triangulation
that is most convenient for studying the action of braiding motions. The
notation we use for counting crossings is illustrated in
Figure~\ref{fig:triangulation}. For $n$ punctures the resulting encoding
$\{x_i, y_i, z_i\}$ contains $3n$ crossing numbers.

To ensure that each homotopic lamination produces the same set of crossings we
insist that the lamination is first pulled tight, which means that no loops
are allowed where the lamination enters and leaves a triangle by the same
edge. Under the pulled-tight assumption, the set of crossing numbers uniquely
identifies the lamination. The reason for this is that the path of the
lamination can be determined uniquely in each triangle, and therefore the
global solution is found to be unique by patching the triangles together. The
unconvinced reader is invited to attempt to find any other solution for the
given crossing numbers that is not homotopic to the lamination shown in
Figure~\ref{fig:puzzle}.

\begin{figure}
\begin{center}
\includegraphics{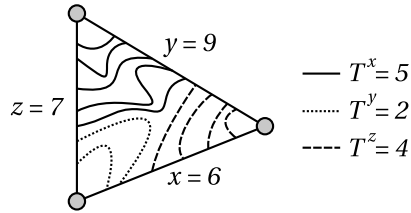}
\end{center}
\caption{The lamination inside a triangle~$T$ (where~$T$ is~$U_i$ or~$L_i$
  from Figure~\ref{fig:triangulation}) can be completely reconstructed from
  the set of crossing numbers $\{x_i,y_i,z_i\}$. To construct the solution,
  each triangle is solved separately, and the triangles are patched together
  in the obvious way. The solution in a single triangle is
  illustrated. Equation \eqref{eq:txyz} determines the number of curves that
  must pass directly between each pair of edges. Up to homotopy, there is only
  one way to draw these curves without crossings.}
\label{fig:triangle}
\end{figure}

To draw the lamination from the set of crossing numbers we explain the
solution procedure for each triangle. Consider the triangle~$T$ depicted in
Figure~\ref{fig:triangle}, where~$T$ is~$U_i$ or~$L_i$ from
Figure~\ref{fig:triangulation}.  The crossing numbers with each of the three
edges are $x$, $y$ and $z$.  In our notation, any part of the lamination
passing between edges $x$ and $y$ is counted by $T^z$, and, likewise, $T^y$
counts lines passing between $x$ and $z$ and $T^x$ counts lines between $y$
and $z$. Since we assume the lamination is pulled tight at all times, there
cannot be any lines that enter and leave by the same edge, so $T^x+T^y+T^z$
counts all the lines. A simple bit of bookkeeping shows we must have
$x=T^y+T^z$, $y=T^x+T^z$ and $z=T^x+T^y$, which gives
\begin{equation}
\begin{split}
T^x &= \tsfrac12 \left( y+z-x \right),\\
T^y &= \tsfrac12 \left( x+z-y \right),\\
T^z &= \tsfrac12 \left( x+y-z \right).
\end{split}
\label{eq:txyz}
\end{equation}
Given $T^x$, $T^y$ and $T^z$, there is then only one way to draw the lines
without them crossing each other. The $T^x$, $T^y$ and $T^z$ lines are
highlighted using different dash patterns for the example in
Figure~\ref{fig:triangle}. The numbers given by \eqref{eq:txyz}
are crucial to the arguments that follow.

\section{Deformation of laminations under braid operations}
\label{sec:deformation}

In this section we describe the effect of braiding operations on a
lamination. We start with a lamination that is pulled tight, and we record the
initial set of crossing numbers $\{x_i,y_i,z_i\}$. All we have to do now is to
determine how these numbers are updated under the action of each braid
operation. 

To calculate the new set of crossings we simply determine the number of
crossings with the {\it preimage} of each edge. We denote the preimage of edge
$e$ by $e^*$, which is a curve that becomes $e$ (up to homotopy) after the
braid operation. Our argument is that the number of crossings of $e^*$ before
the braid operation has to be equal to the number of crossings with $e$
afterwards.  This must be the case, as the only way the number of crossings
could change is if the end of a loop were to cross through the edge as it is
deformed from $e^*$ to $e$ --- but by the pulled-tight assumption any such
loop would have to be wrapped around a puncture, and by construction no
punctures pass through the edge as it deforms from $e^*$ to $e$.

It will turn out that we only have to consider explicitly the braid group
operations $\rho_i$, $\rho_i^{-1}$ $(i=1,\ldots,n)$ and $\sigma_i$,
$\sigma_i^{-1}$ $(i=1,\ldots,n-1)$.  We then determine the effect of other
independent group elements $\tau_i$ and $\tau_i^{-1}$ by invoking group
presentation rules and the operation $\sigma_n$ (see
Figure~\ref{fig:bsd}(b)).  This is explained in detail in
Section~\ref{sec:xingtaui} for those not familiar with the braid group
presentation.  For convenience we assume in what follows that indices are
treated `modulo' $n$, so that the puncture to the right of puncture $n$ is
puncture $1$.

\subsection{Crossing update rules for $\boldsymbol{\rho_i}$}
\label{sec:xingrhoi}

\begin{figure}
\begin{center}
\includegraphics{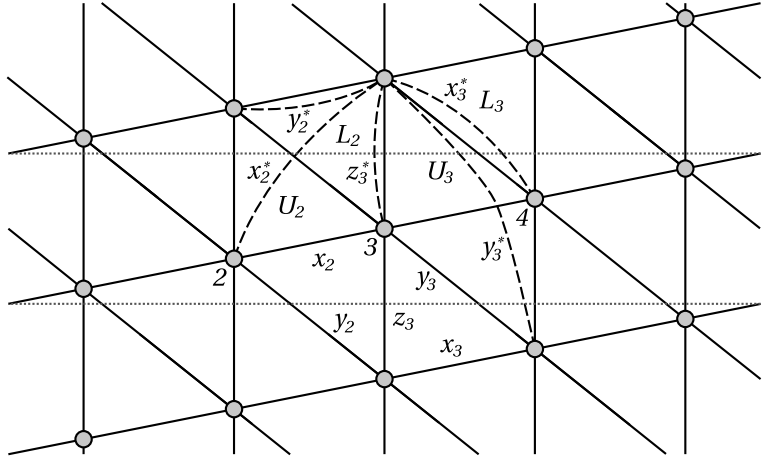}
\end{center}
\caption{Preimage illustration for $\rho_3$. Only edges that
are incident with the moving puncture can have a change in crossing number.
For $\rho_3$ the five edges are $x_2$, $y_2$, $x_3$, $y_3$ and $z_3$.  In this
example $x_2$ and $y_3$ are the exact preimages of $y_2$ and $x_3$,
respectively. The edge $z_3$ is the preimage of itself, so in fact its
crossing number will not change under $\rho_3$. The preimages of $x_2$ and
$y_3$ are not edges in the triangulation, so more work is required to
determine how many crossings are made, as shown in
Figure~\ref{fig:rhopuzzle}.}
\label{fig:rho}
\end{figure}

For the braid operation $\rho_i$, we need only consider edges incident on the 
$i$th puncture, since the number of crossings with other edges will remain 
unchanged. The relevant preimages are shown in Figure~\ref{fig:rho} as dashed 
lines.
Many of the preimages are other edges (e.g.\ $x_{i-1}$ is the preimage
$y_{i-1}^*$), so the number of crossings is already known. This is not a
coincidence: the triangulation was chosen for this property.
\begin{figure}
\begin{center}
\includegraphics{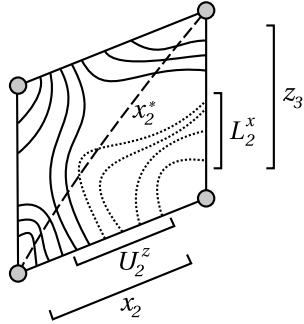}
\end{center}
\caption{The number of crossings with a preimage that is not part of
the triangulation may be found by simple bookkeeping. We illustrate here
how to determine the number of crossings with the $x_2^*$ preimage. All 
curves entering $x_2$ and $z_3$ must cross $x_2^*$ (by the pulled-tight
assumption), unless they loop directly from $x_2$ to $z_3$ (shown dotted). 
The number of these loops is exactly $\min(U_2^z,L_2^x)$. Hence the number
of preimage crossings is $x_2+z_3-2\min(U_2^z,L_2^x)$. Note that curves
that cross over the preimage and immediately back again are automatically
discounted, so that the pulled-tight assumption still holds for the updated
set of crossing numbers.} 
\label{fig:rhopuzzle}
\end{figure}
The two preimages $x_{i-1}^*$ and $y_i^*$ require a little more
attention. In Figure~\ref{fig:rhopuzzle} we illustrate that the number of
crossings $x_{i-1}^*$ can be calculated by a simple bookkeeping argument. The
number of crossings $y_i^*$ can be deduced in a similar way.

The update rules for all the edges considered are
\begin{equation}
\begin{split}
x_{i-1}^* &= x_{i-1} + z_i - 2 \min(U_{i-1}^z,L_{i-1}^x)
= x_{i-1} + z_{i-1} - 2 \min(U_{i-1}^x,L_{i-1}^z),\\ 
y_{i-1}^* &= x_{i-1}, \\
x_i^*     &= y_i,\\
y_i^*     &= z_i + y_i - 2 \min(U_{i}^y,L_{i}^z)
= y_i + z_{i+1}  - 2 \min(U_{i}^z,L_{i}^y), \\
z_i^*     &= z_i.
\end{split}
\label{eq:rhoup}
\end{equation}

\subsection{Crossing update rules for $\boldsymbol{\rho_i^{-1}}$}
\label{sec:xingrhoii}

We can work out the update rule for $\rho_i^{-1}$ by noting the
$\pi$-rotational symmetry of the triangulation about a puncture and
relabelling the variables in the rules \eqref{eq:rhoup} given above. The
update rules for affected crossing numbers are
\begin{equation}
\begin{split}
x_{i-1}^* &= y_{i-1}, \\ 
y_{i-1}^* &= y_{i-1} + z_i - 2 \min(U_{i-1}^z,L_{i-1}^y) 
= y_{i-1} + z_{i-1} - 2 \min(U_{i-1}^y,L_{i-1}^z),  \\
x_i^*     &= z_i + x_i - 2 \min(U_{i}^x,L_{i}^z) 
= x_i + z_{i+1} - 2 \min(U_{i}^z,L_{i}^x), \\
y_i^*     &= x_i, \\
z_i^*     &= z_i.
\end{split}
\label{eq:rhoiup}
\end{equation}

\subsection{Crossing update rules for $\boldsymbol{\sigma_i}$}
\label{sec:xingsigmai}

\begin{figure}
\begin{center}
\includegraphics{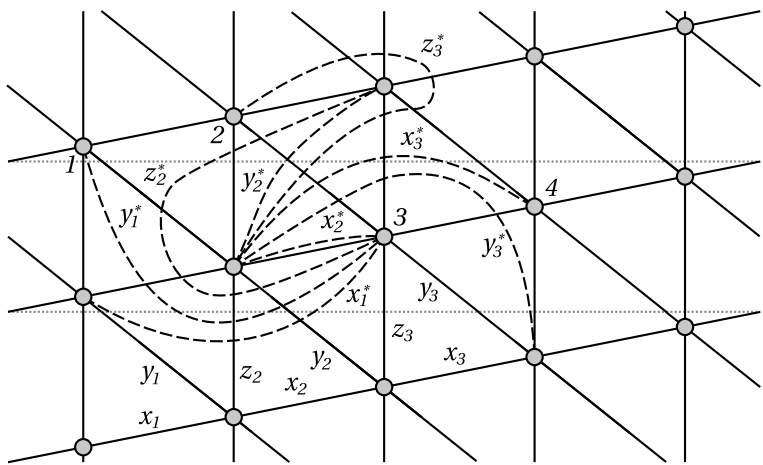}
\end{center}
\caption{Preimage illustration for $\sigma_2$. The dashed lines show the
preimages of all edges whose crossing number may change.  Edge $x_2$ is its
own preimage, but all other preimages require some bookkeeping.
Figure~\ref{fig:sigmapuzzle} illustrates how to compute the crossings with
$x_1^*$ and $y_1^*$. Computation of $y_2^*$ is done as in
Figure~\ref{fig:rhopuzzle}. Preimages $x_3^*$ and $y_3^*$ are handled in the
same way as $x_1^*$ and $y_1^*$, by invoking rotational symmetry. The
difficult preimage problems are $z_2^*$ and $z_3^*$, as these preimages cross
through seven triangles; the solution is described in
Section~\ref{sec:xingsigmai}.}
\label{fig:sigma}
\end{figure}

The same ideas can be used to determine the updated crossing numbers following
a $\sigma_i$ operation. However, since there are two moving punctures, more
edges are affected, and the preimages are therefore more complicated. An
example of a preimage diagram for $\sigma_2$ is shown in
Figure~\ref{fig:sigma}. In general, for $\sigma_i$ the affected crossing
numbers are $x_{i-1}$, $y_{i-1}$, $x_{i}$, $y_{i}$, $z_{i}$, $x_{i+1}$,
$y_{i+1}$ and $z_{i+1}$. This includes the operation $\sigma_n$ which switches
the first and last punctures across the periodic boundaries (see
Figure~\ref{fig:bsd}(b)); we shall require this operation in
Sections~\ref{sec:xingtaui}--\ref{sec:xingtauii} to find the update rules for
$\tau_i$ and $\tau_i^{-1}$.

\begin{figure}
\begin{center}
\includegraphics{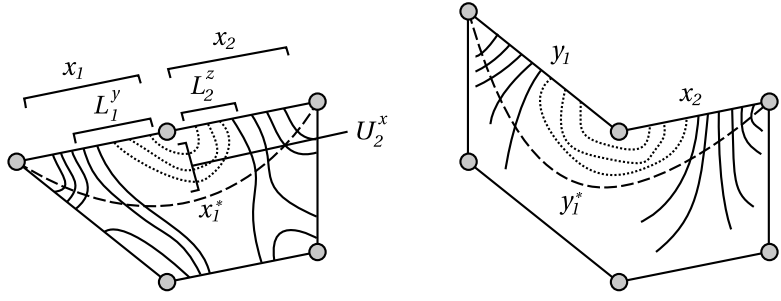}
\end{center}
\caption{Two slightly more difficult preimage puzzles for the operation
$\sigma_2$. In a similar way to in Figure~\ref{fig:rhopuzzle}, the number of
crossings with preimage $x_1^*$ is given by the number of curves crossing
$x_1$ and $x_2$, minus twice the number of loops directly between $x_1$ and
$x_2$. The number of such loops is exactly $\min(L_1^y,U_2^x,L_2^z)$. Hence
$x_1^* = x_1 + x_2 - 2 \min(L_1^y,U_2^x,L_2^z)$. The preimage problem for
$y_1^*$ is similar, but involves four triangles and hence a minimum of four
numbers, so that $y_1^* = y_1 + x_2 - 2 \min(U_1^z,L_1^y,U_2^x,L_2^z)$.}
\label{fig:sigmapuzzle}
\end{figure}

The edge $x_i$ is its own preimage, and the number of crossings with $y_i^*$
is determined using the quadrilateral trick illustrated in
Figure~\ref{fig:rhopuzzle}. The number of crossings with $x_{i-1}^*$,
$y_{i-1}^*$, $x_{i+1}^*$ and $y_{i+1}^*$ can be found by extending the
quadrilateral solution over three and four triangles, as illustrated in
Figure~\ref{fig:sigmapuzzle}.

\begin{figure}
\begin{center}
\includegraphics{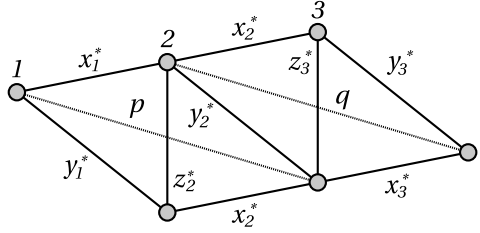}
\end{center}
\caption{The number of crossings with $z_i^*$ and $z_{i+1}^*$ are difficult to
determine directly from the old set of crossing numbers. However they may be
calculated using the quadrilateral solution shown in
Figure~\ref{fig:rhopuzzle} from the already updated crossing numbers.}
\label{fig:sigmapuzzle2}
\end{figure}

The two remaining crossing numbers to find are $z_i^*$ and $z_{i+1}^*$. Since
these preimages pass through a total of seven triangles it is non-trivial to
determine the number of crossings directly. Instead we employ a trick and
deduce $z_i^*$ and $z_{i+1}^*$ by invoking the quadrilateral solution again
with the {\it updated} crossing numbers. We introduce temporary edges $p$ and
$q$ directly between punctures $i-1$ and $i+1$ and between $i$ and $i+2$, as
shown in Figure~\ref{fig:sigmapuzzle2}. The preimage of $p$ is $y_{i-1}$, and
the preimage of $q$ is $y_{i+1}$. Since $x_{i-1}^*$, $y_{i-1}^*$, $p$,
$x_i^*$, $y_i^*$, $q$, $x_{i+1}^*$ and $y_{i+1}^*$ are known, $z_i^*$ and
$z_{i+1}^*$ can be deduced easily.

The collected update rules for the affected crossing numbers are
\begin{equation}
\begin{split}
x_{i-1}^* &= x_{i-1} + x_i - 2 \min(L_{i-1}^y,U_{i}^x,L_i^z), \\
y_{i-1}^* &= y_{i-1} + x_i - 2 \min(U_{i-1}^z,L_{i-1}^y,U_i^x,L_i^z),\\
x_i^*     &= x_i,\\
y_i^*     &= z_i + x_i - 2 \min(U_i^x,L_i^z) 
= x_i + z_{i+1} - 2 \min(U_i^z,L_i^x), \\
z_i^*     &= x_{i-1}^* + y_{i-1}^* 
- \min(y_{i-1}^*+y_{i-1}-x_i^*,x_{i-1}^*+y_{i-1}-y_i^*), \\
x_{i+1}^* &= x_i + x_{i+1} - 2 \min(U_i^z,L_i^x,U_{i+1}^y), \\ 
y_{i+1}^* &= x_i + y_{i+1} - 2 \min(U_i^z,L_i^x,U_{i+1}^y,L_{i+1}^z), \\
z_{i+1}^* &= x_i^* + y_i^* 
- \min(x_i^*+y_{i+1}-y_{i+1}^*,y_i^*+y_{i+1}-x_{i+1}^*).
\end{split}
\label{eq:sigmaup}
\end{equation}

\subsection{Crossing update rules for $\boldsymbol{\sigma_i^{-1}}$}
\label{sec:xingsigmaii}

\begin{figure}
\begin{center}
\includegraphics{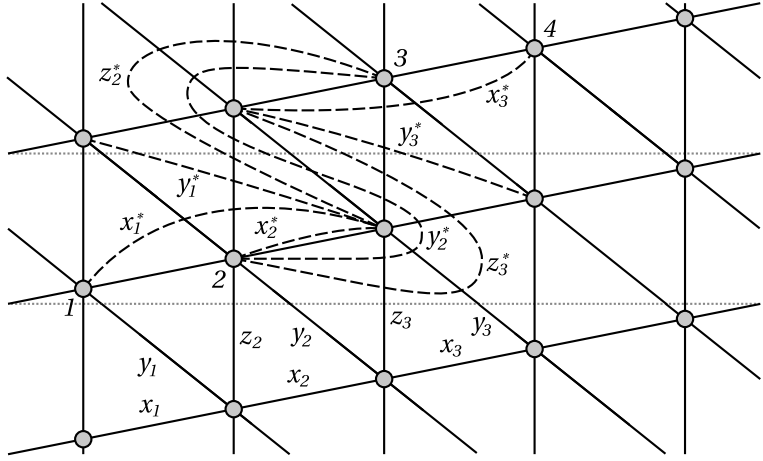}
\end{center}
\caption{Preimage illustration for $\sigma_2^{-1}$. Note that due to
asymmetry the preimage problems are slightly different to those for
$\sigma_2$.}
\label{fig:sigmai}
\end{figure}

Since the triangulation does not have a reflection symmetry about a vertical
line through the midpoint of two punctures, it is not possible to deduce the
update rules for $\sigma_i^{-1}$ by a relabelling in the rules for $\sigma_i$.
An example of a preimage diagram for $\sigma_2^{-1}$ is shown in
Figure~\ref{fig:sigmai}. The preimage curve $y_i^*$ is the most complicated
yet, as it passes through ten triangles. However the crossing number of all
these preimages can be calculated as before using the techniques illustrated
in Figures \ref{fig:rhopuzzle}, \ref{fig:sigmapuzzle} and
\ref{fig:sigmapuzzle2}.

The update rules for the affected crossing numbers are
\begin{equation}
\begin{split}
x_{i-1}^* &= x_{i-1} + x_i - 2 \min(U_{i-1}^z,L_{i-1}^x,U_i^y), \\
y_{i-1}^* &= y_{i-1} + x_i - 2 \min(L_{i-1}^x,U_i^y) 
= x_{i-1} + y_i - 2 \min(L_{i-1}^y,U_i^x), \\
x_i^*     &= x_i, \\
y_i^*     &= x_i^* + z_i^* - \min(z_i^*+y_i-x_i^*,x_i^*+y_i-z_{i+1}^*), \\
z_i^*     &= x_i + y_i - 2 \min(U_i^x,L_{i-1}^y,U_{i-1}^z,L_{i-1}^x,U_i^y), \\
x_{i+1}^* &= x_i + x_{i+1} - 2 \min(L_i^y,U_{i+1}^x,L_{i+1}^z), \\ 
y_{i+1}^* &= y_i + x_{i+1} 
- 2 \min(L_i^x,U_{i+1}^y) = x_i + y_{i+1} - 2 \min(L_i^y,U_{i+1}^x), \\
z_{i+1}^* &= x_i + y_i - 2 \min(L_i^y,U_{i+1}^x,L_{i+1}^z,U_{i+1}^y,L_i^x). 
\end{split}
\label{eq:sigmaiup}
\end{equation}

\subsection{Crossing update rules for $\boldsymbol{\tau_i}$}
\label{sec:xingtaui}

In Sections~\ref{sec:xingrhoi}--\ref{sec:xingsigmaii} we showed how to update
the set of crossing numbers $\{x_i,y_i,z_i\}$ for the braid operations
$\rho_i$, $\rho_i^{-1}$, $\sigma_i$ and $\sigma_i^{-1}$.  To complete the set
of update rules for any braid we must give the corresponding rules for
$\tau_i$ and $\tau_i^{-1}$. In performing $\tau_i$ the $i$th puncture moves
once around the torus in the horizontal direction (see
Figure~\ref{fig:bsd}). In doing so it passes through many edges in our
triangulation, so it is difficult to draw the preimages and to derive the
number of crossings directly.
\begin{figure}
\begin{center}
\includegraphics{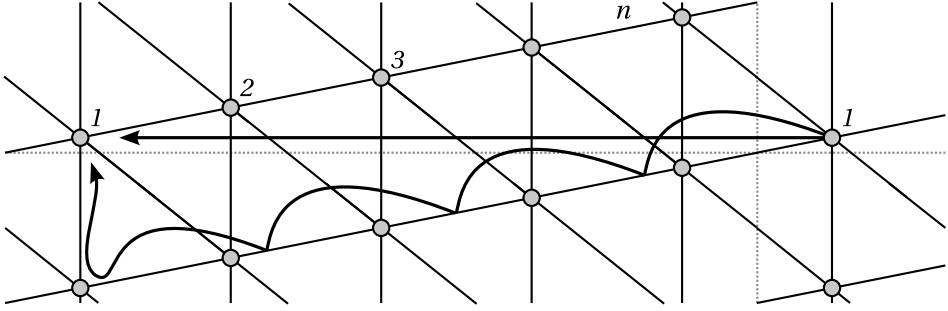}
\end{center}
\caption{The operation $\tau_1$ is achieved using a sequence of
$\sigma_j^{-1}$ operations, including the operation $\sigma_n^{-1}$ (see
Figure~\ref{fig:bsd}), followed by a single $\rho_1^{-1}$. This is described
in more detail in the text.}
\label{fig:tau}
\end{figure}
However, we can deduce the update rules for $\tau_i$ and $\tau_i^{-1}$ by
appealing to group properties that relate $\sigma_i$, $\rho_i$ and $\tau_i$.
In Figure~\ref{fig:tau} we illustrate how $\tau_1$ is achieved through a
sequence of $\sigma_i^{-1}$ (including $\sigma_n^{-1}$ defined in
Figure~\ref{fig:bsd}(b)) followed by one $\rho_1^{-1}$. The other $\tau_i$ and
$\tau_i^{-1}$ are produced in a similar manner. A computational recipe for
each~$\tau_i$ operation is given in this section, and~$\tau_i^{-1}$ in the
following section.

To calculate the updated set of crossing numbers $\{x_i,y_i,z_i\}$ for
$\tau_i$ do the following:
\begin{enumerate}
\item Use equation \eqref{eq:sigmaiup} to perform, in turn,
$\sigma_{i-1}^{-1}$, $\sigma_{i-2}^{-1}$, $\ldots$, $\sigma_{i+2}^{-1}$ and
$\sigma_{i+1}^{-1}$. Treat the indices `modulo' $n$, so that $\sigma_n^{-1}$
follows $\sigma_1^{-1}$.
\item Relabel $x_i \leftarrow x_{i+1}$, $y_i \leftarrow y_{i+1}$ and $z_i
\leftarrow z_{i+1}$. This leaves all punctures except the $i$th one in the
correct position.
\item Use equation \eqref{eq:rhoiup} to perform $\rho_i^{-1}$.
\end{enumerate}

\subsection{Crossing update rules for $\boldsymbol{\tau_i^{-1}}$}
\label{sec:xingtauii}

To calculate the updated set of crossing numbers $\{x_i,y_i,z_i\}$ for
$\tau_i^{-1}$ invert the operation of Section~\ref{sec:xingtaui} as follows:
\begin{enumerate}
\item Use equation \eqref{eq:rhoup} to perform $\rho_i$.
\item Relabel $x_i \rightarrow x_{i+1}$, $y_i \rightarrow y_{i+1}$ and $z_i
\rightarrow z_{i+1}$. This leaves the punctures in the wrong position, but the
next sequence corrects everything\dots
\item Use equation \eqref{eq:sigmaup} to perform, in
turn, $\sigma_{i+1}$, $\sigma_{i+2}$, $\ldots$, $\sigma_{i-2}$ and
$\sigma_{i-1}$. Treat the indices `modulo' $n$, so that $\sigma_1$ follows
$\sigma_n$.
\end{enumerate}

\section{Illustrations using simple braids}
\label{sec:examples}

\begin{figure}
\begin{center}
\includegraphics[scale=0.35]{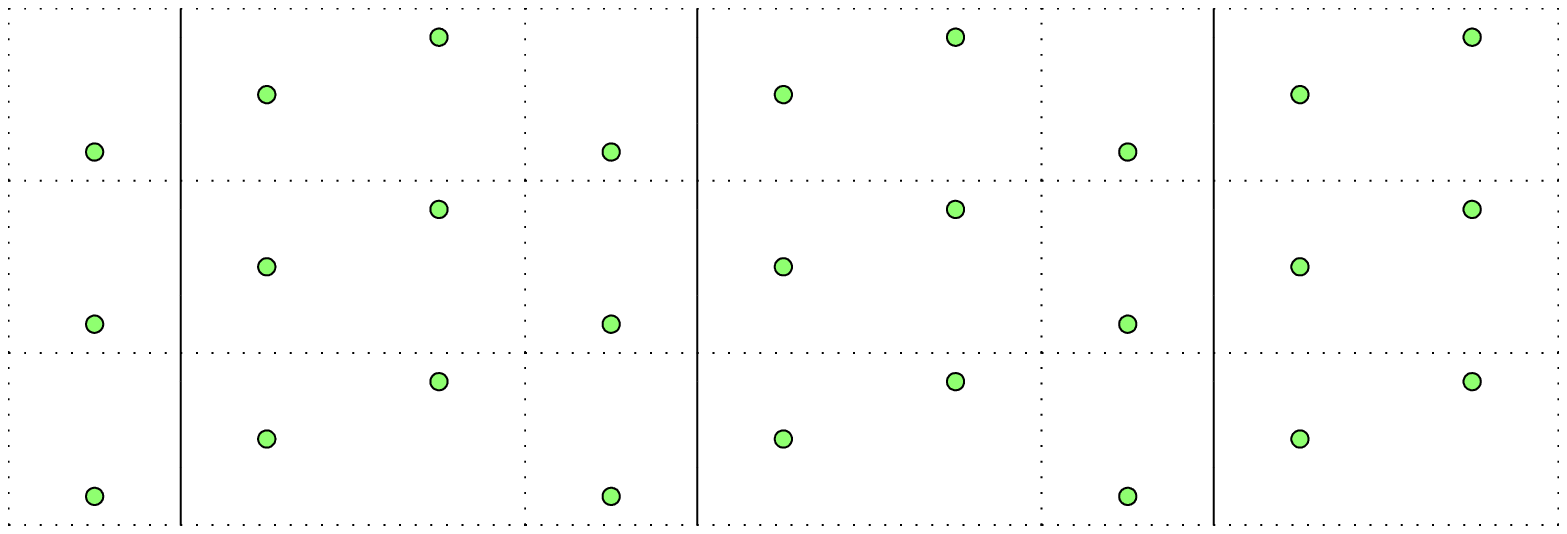} \hspace{0.2cm}
\includegraphics[scale=0.35]{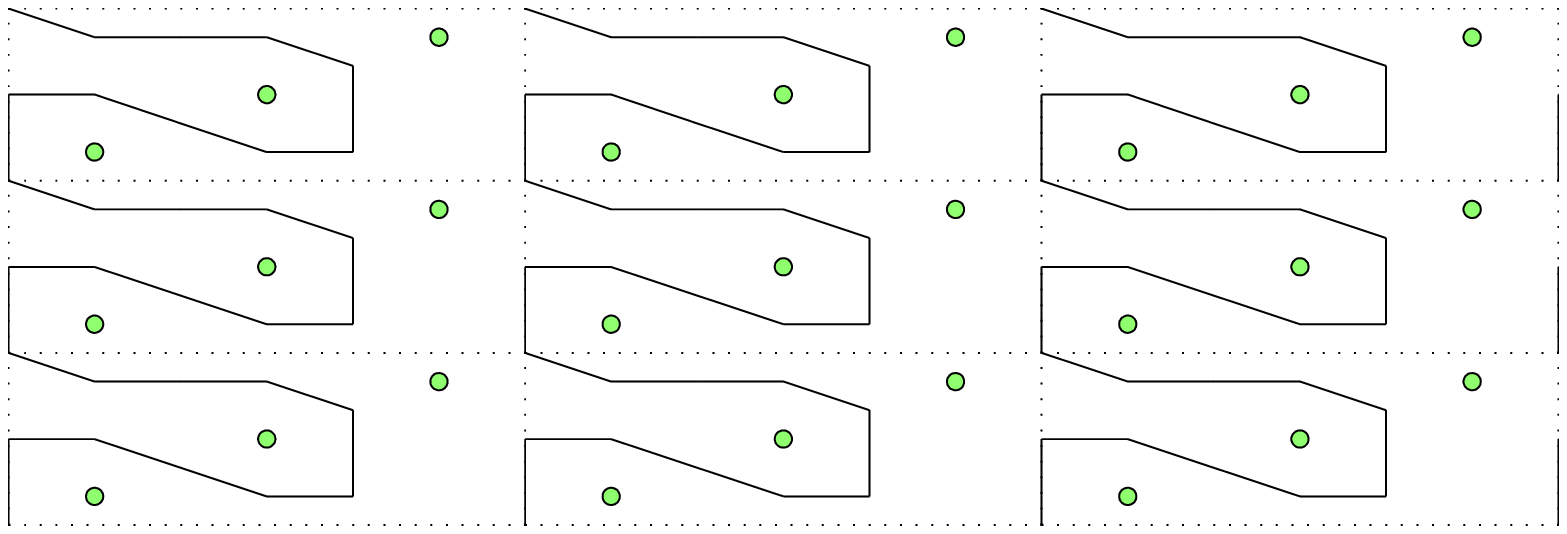}

\vspace{0.2cm}

\includegraphics[scale=0.35]{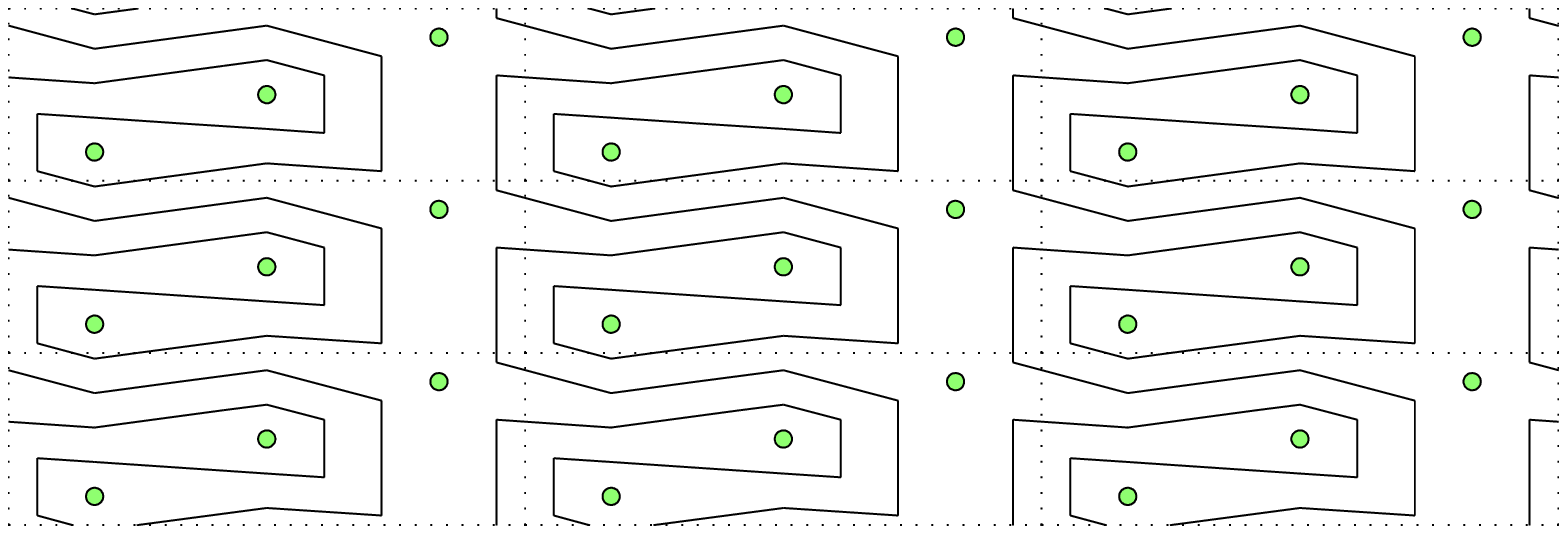} \hspace{0.2cm}
\includegraphics[scale=0.35]{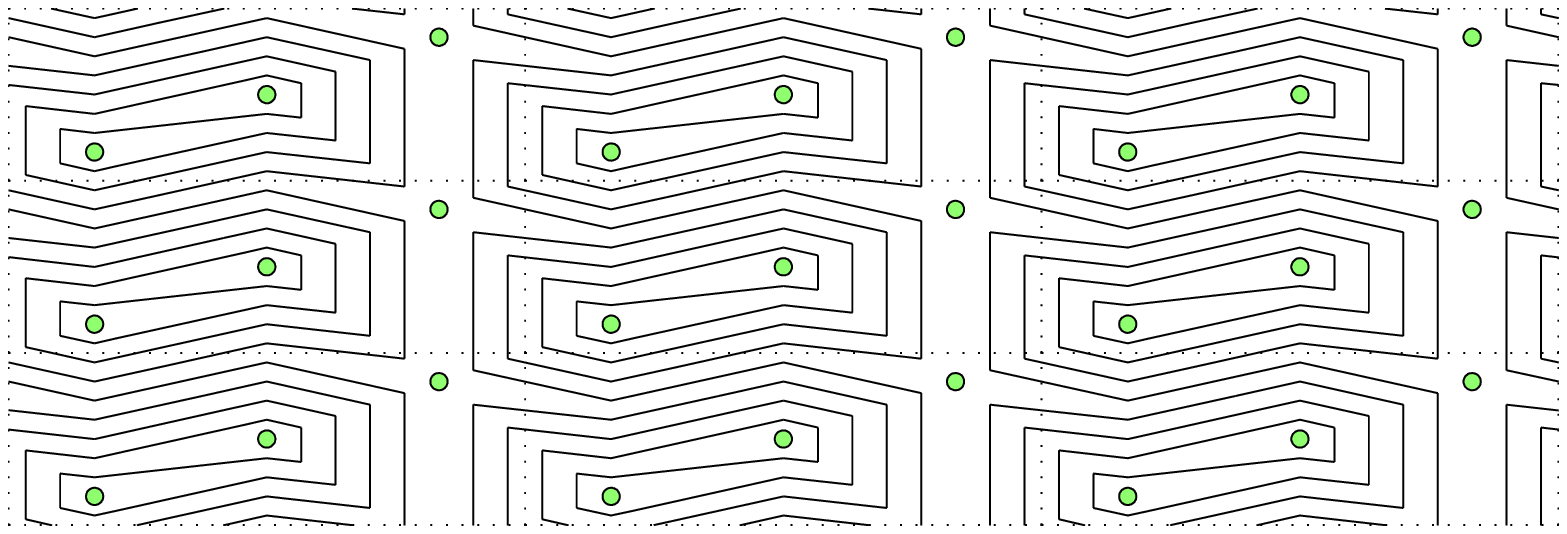}

\vspace{0.2cm}

\includegraphics[scale=0.35]{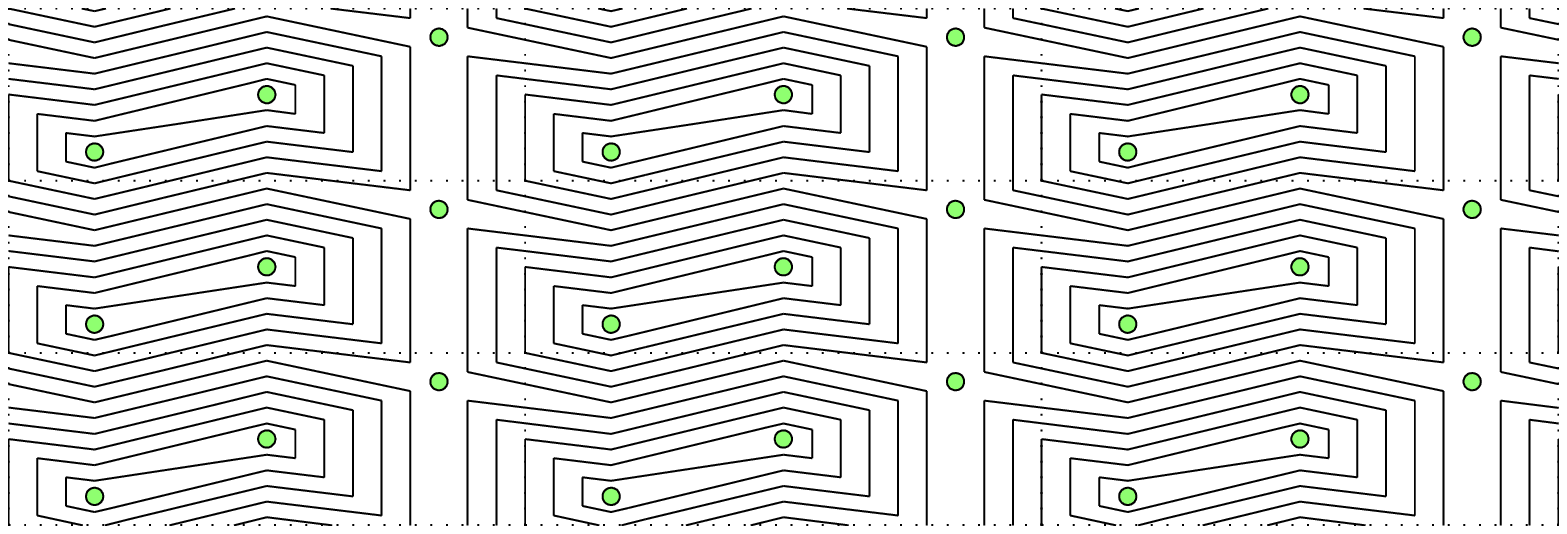} \hspace{0.2cm}
\includegraphics[scale=0.35]{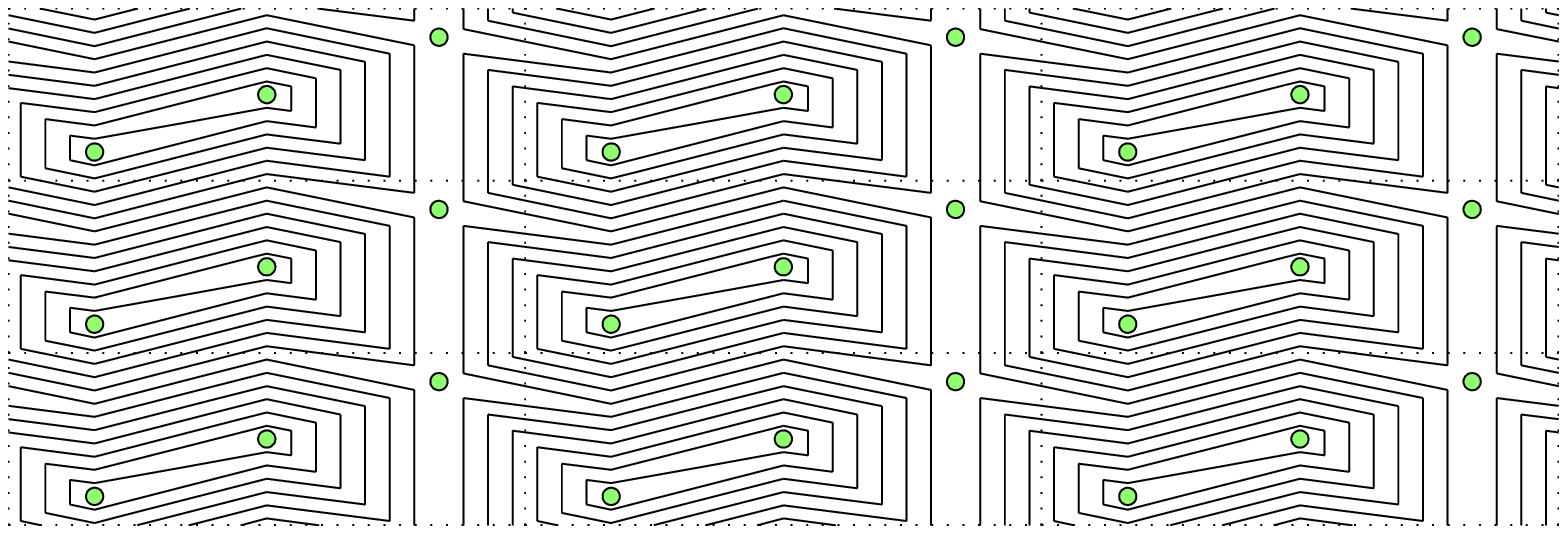}
\end{center}
\caption{Lamination evolution under the braid $\sigma_1$. This results in
linear growth in the number of crossings, and zero entropy.}
\label{fig:sigma1}
\end{figure}

To illustrate the use of the update rules given in
Section~\ref{sec:deformation} we show how a lamination is deformed under some
simple previously studied braids. We have implemented the update rules in a
short C++ program, using the Gnu Multiple Precision library to allow the
number of crossings to grow arbitrarily large whilst maintaining exact
arithmetic. We also have a Matlab script to draw the lamination, one triangle
at a time, using the procedure described in Section \ref{sec:encoding}.  This
was used to produce all the figures in this section.  In each case our initial
lamination is a closed loop that passes between the first two punctures, with
$x_1 = y_1 = 1$ and all other $x_i$, $y_i$ and $z_i$ set to zero (see the
upper-left frame in Figure~\ref{fig:sigma1}).

Figure~\ref{fig:sigma1} shows the roll up of the lamination under the repeated
action of the planar braid $\sigma_1$ with three punctures. In this case it is
clear that the third puncture is redundant. This braid has zero entropy and is
very poor at stirring as it results in linear growth of material lines. Note
that even though coils form around the pair of moving punctures there is always
exactly one crossing of the lamination from one copy of the domain to the copy
above. This is because there was exactly one crossing in the initial
lamination and this is only a planar braid, so it cannot create any further
crossings under the pulled-tight assumption.

\begin{figure}
\begin{center}
\includegraphics[scale=0.35]{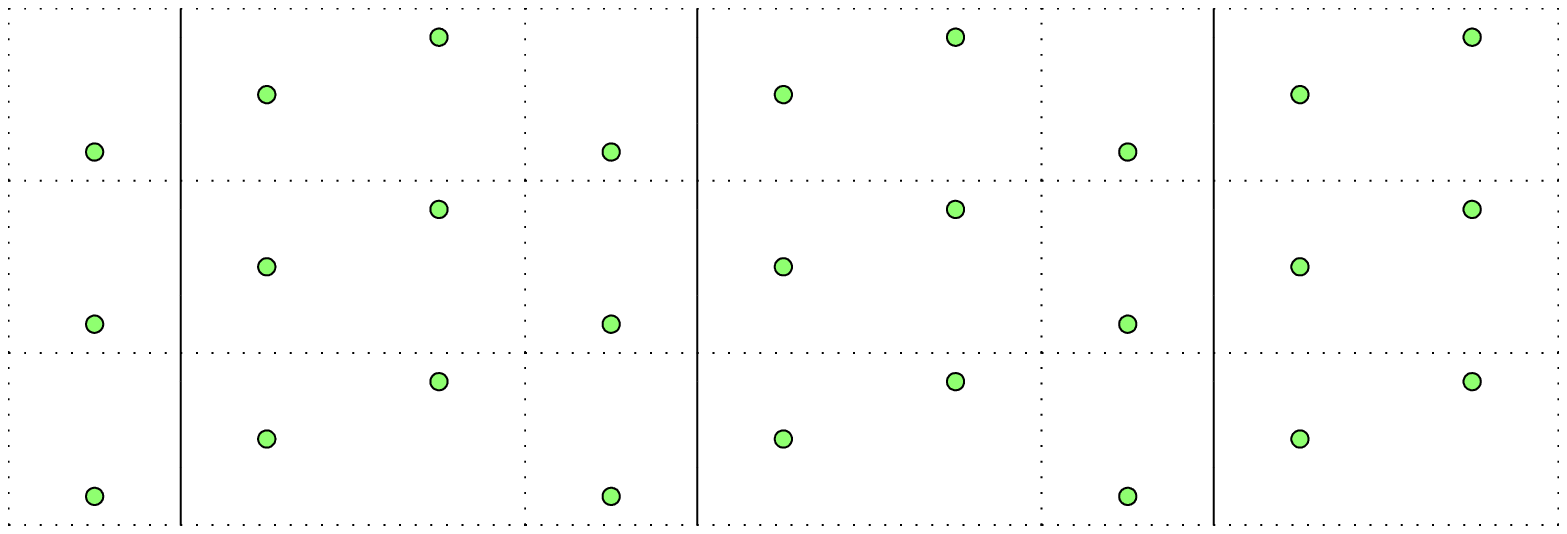} \hspace{0.2cm}
\includegraphics[scale=0.35]{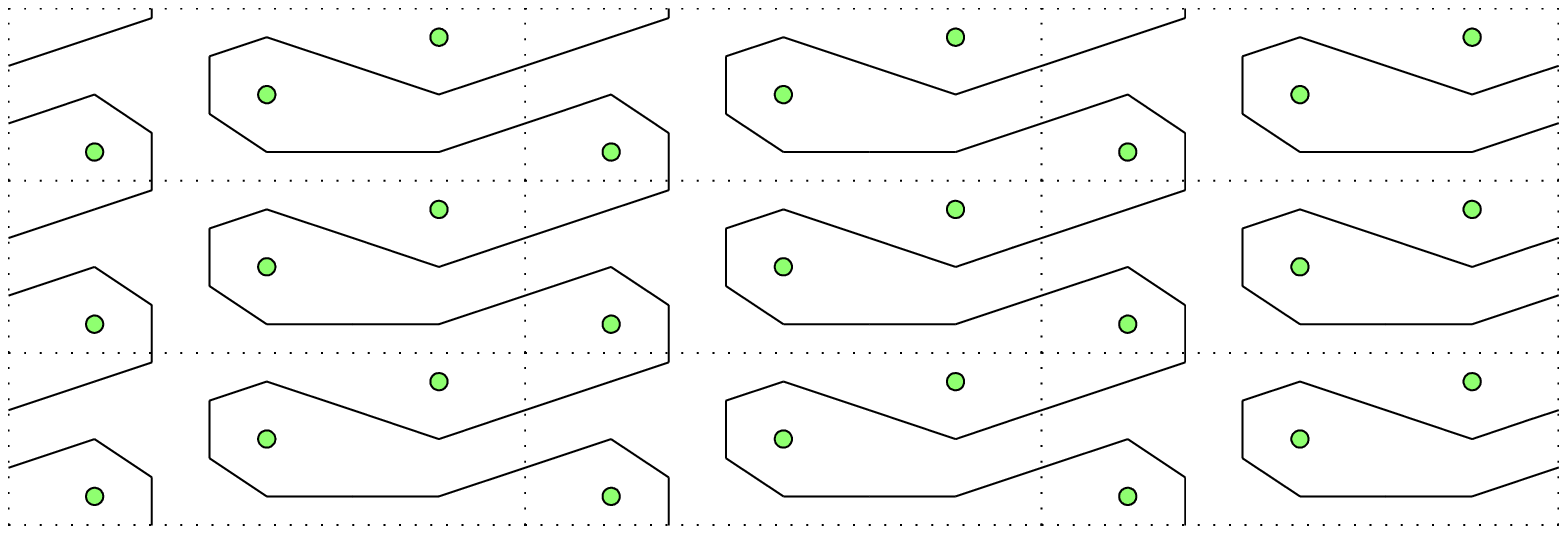}

\vspace{0.2cm}

\includegraphics[scale=0.35]{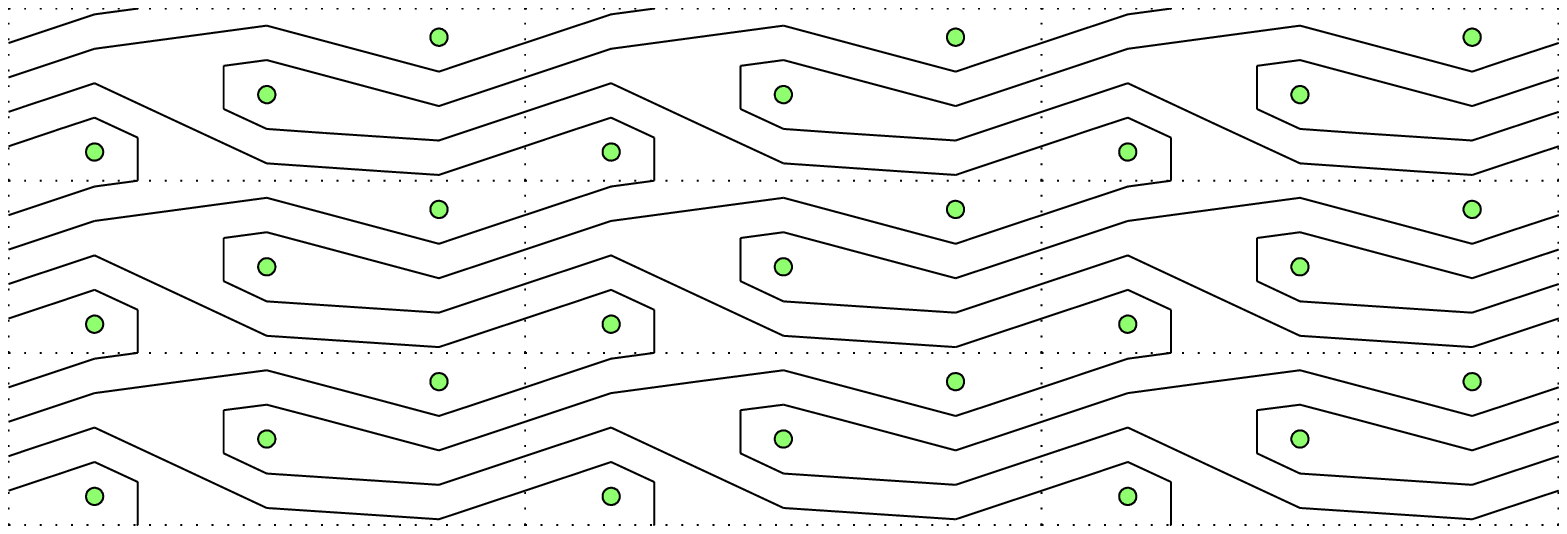} \hspace{0.2cm}
\includegraphics[scale=0.35]{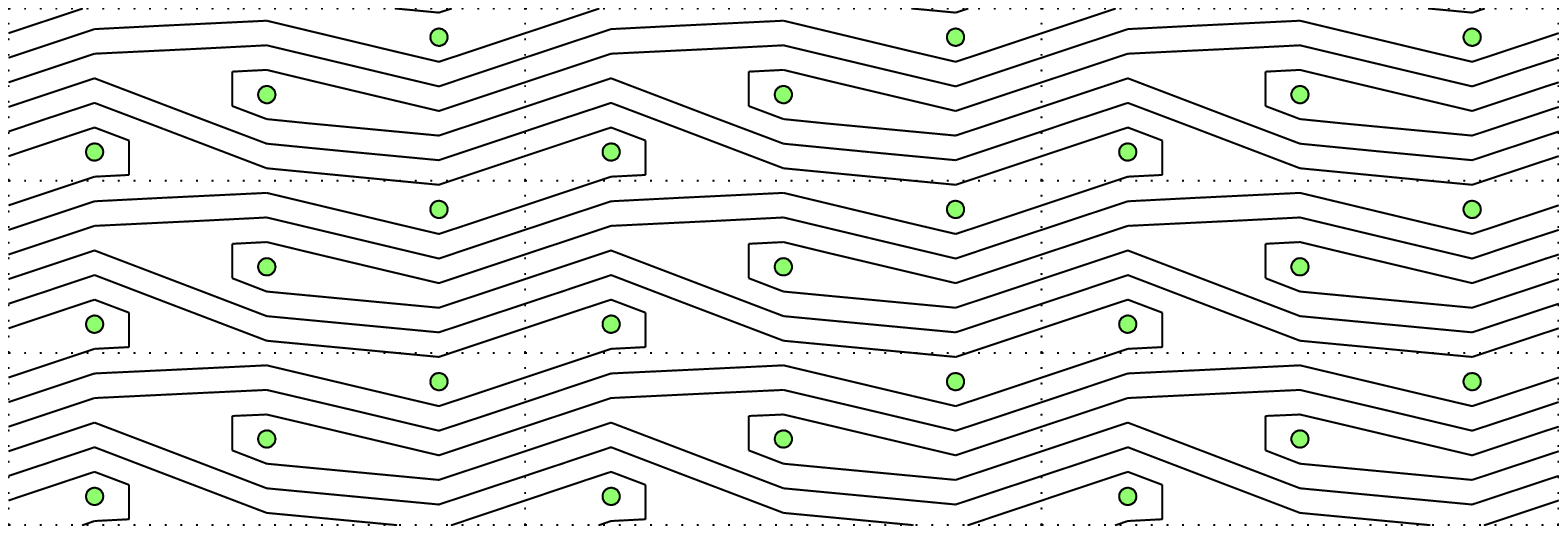}

\vspace{0.2cm}

\includegraphics[scale=0.35]{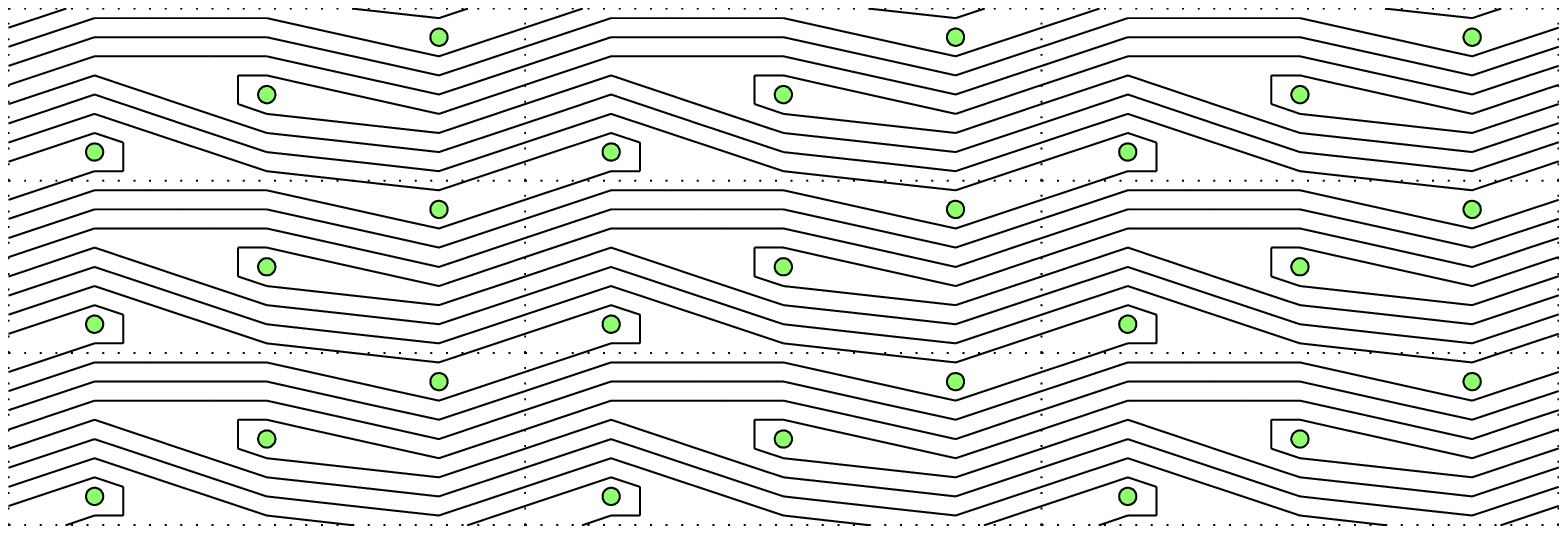} \hspace{0.2cm}
\includegraphics[scale=0.35]{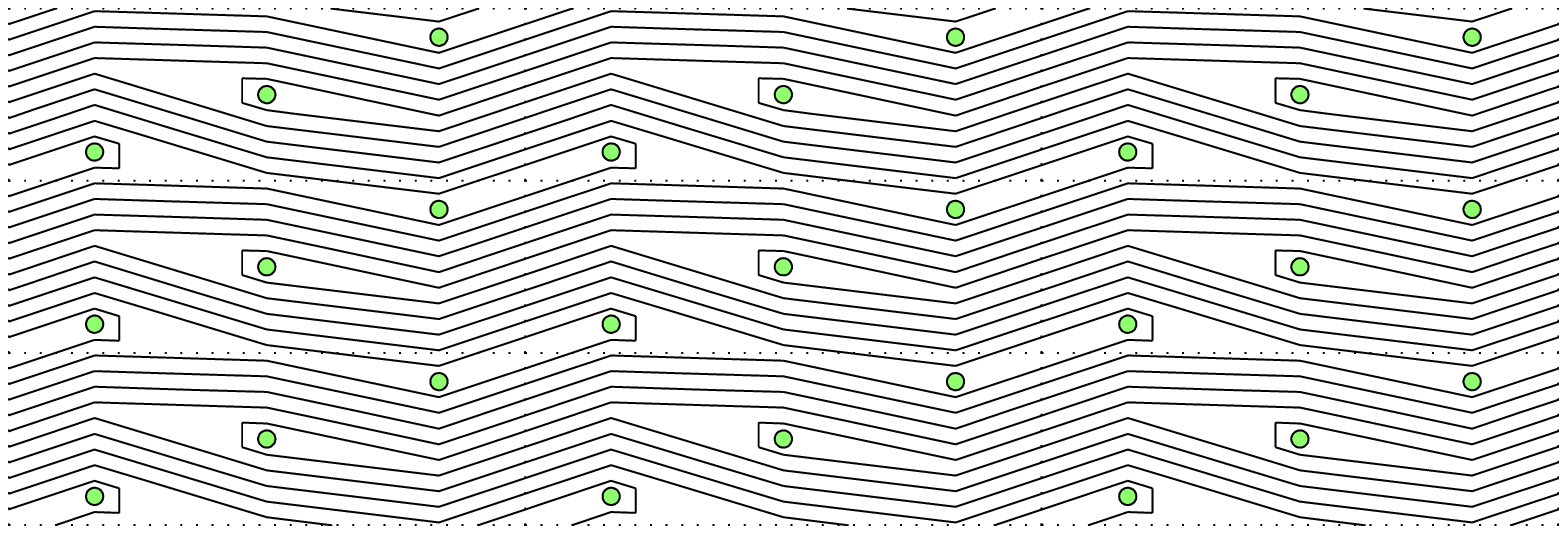}
\end{center}
\caption{Lamination evolution under the braid $\tau_2$. This braid also has
zero topological entropy, but the illustration is provided to verify the
correctness of the procedure for performing $\tau$ using a combination of
$\sigma$ and $\rho$.}
\label{fig:tau2}
\end{figure}

Figure~\ref{fig:tau2} shows the result of repeating $\tau_2$ with three
punctures. In this simple braid the second puncture moves in a straight line
to the left, but catching the lamination on the puncture exactly once. The
illustration is provided to validate the method described in
Sections~\ref{sec:xingtaui}--\ref{sec:xingtauii} for performing $\tau$
operations using a combination of $\sigma$ and $\rho$ motions.

\begin{figure}
\begin{center}
\includegraphics[scale=0.35]{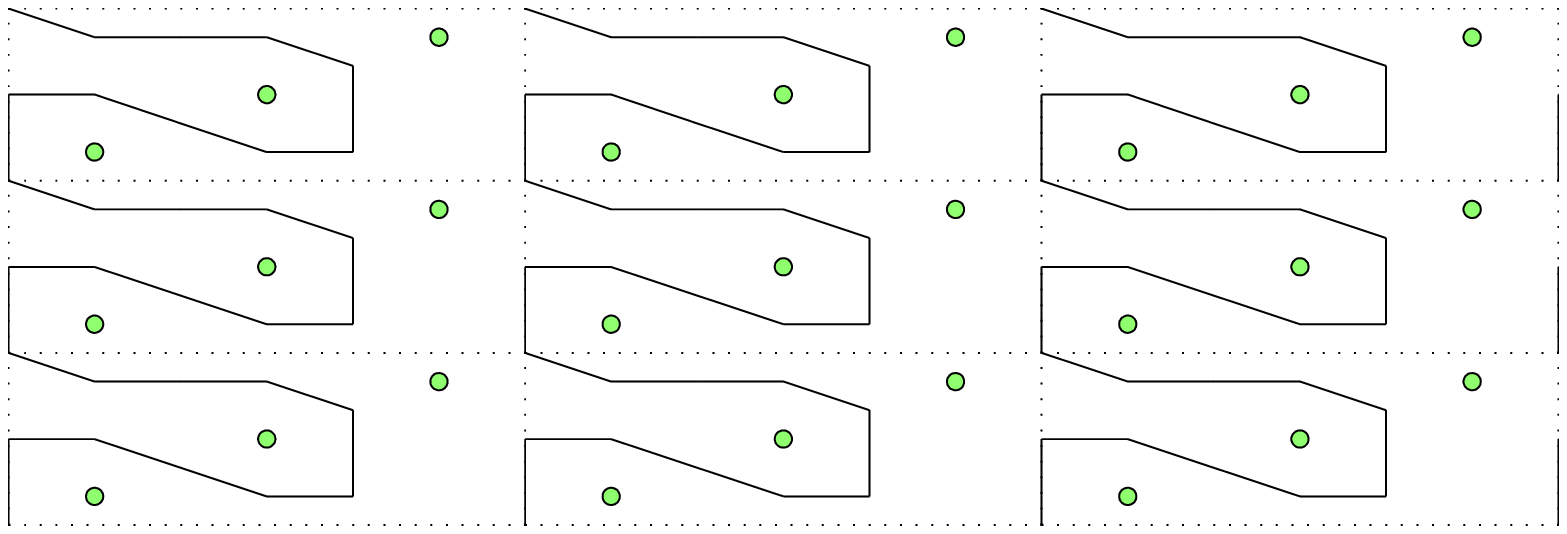} \hspace{0.2cm}
\includegraphics[scale=0.35]{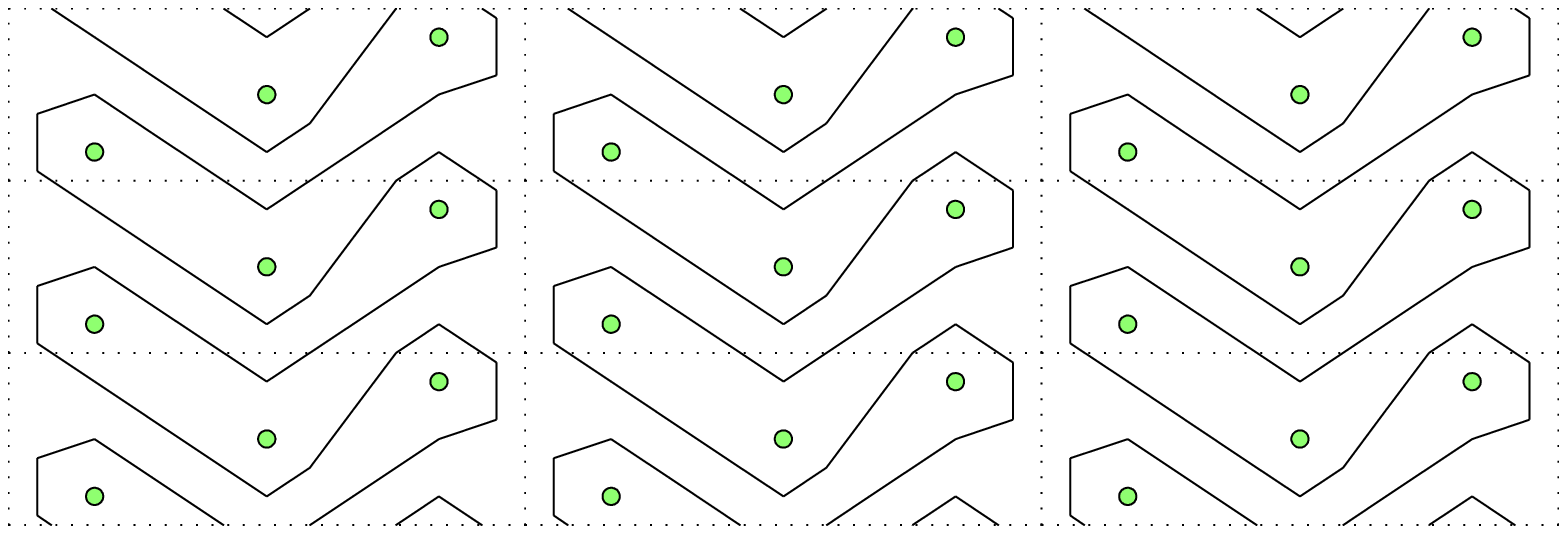}

\vspace{0.2cm}

\includegraphics[scale=0.35]{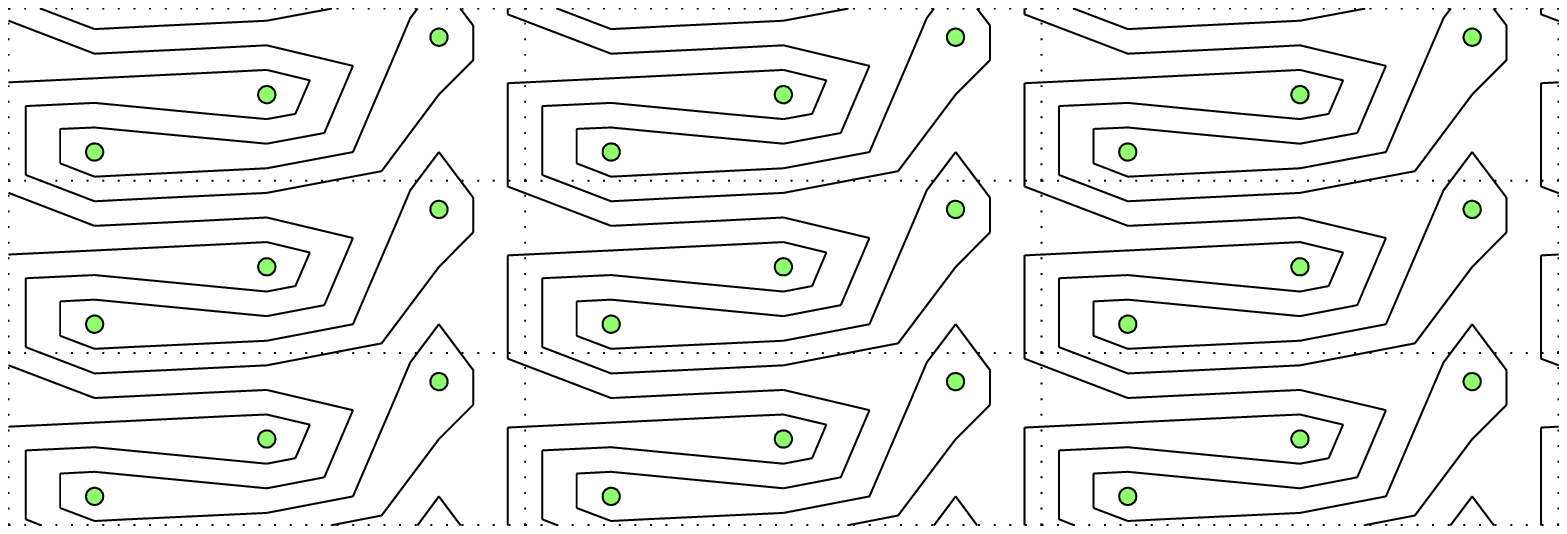} \hspace{0.2cm} 
\includegraphics[scale=0.35]{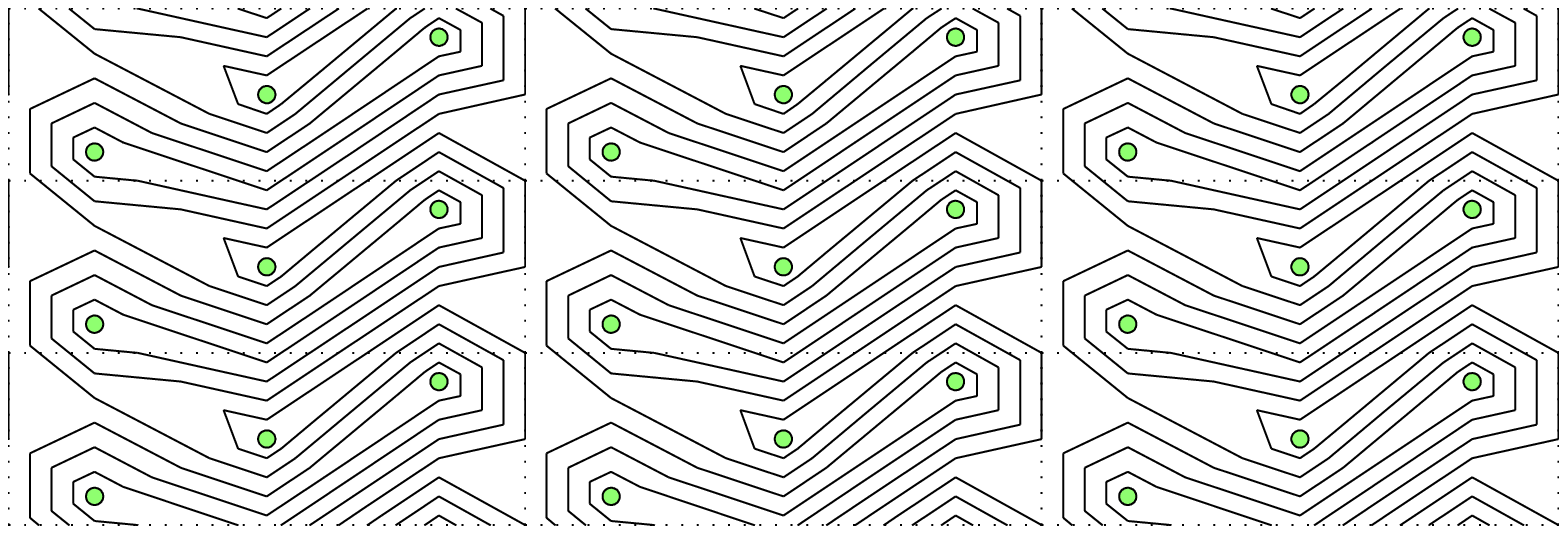}

\vspace{0.2cm}

\includegraphics[scale=0.35]{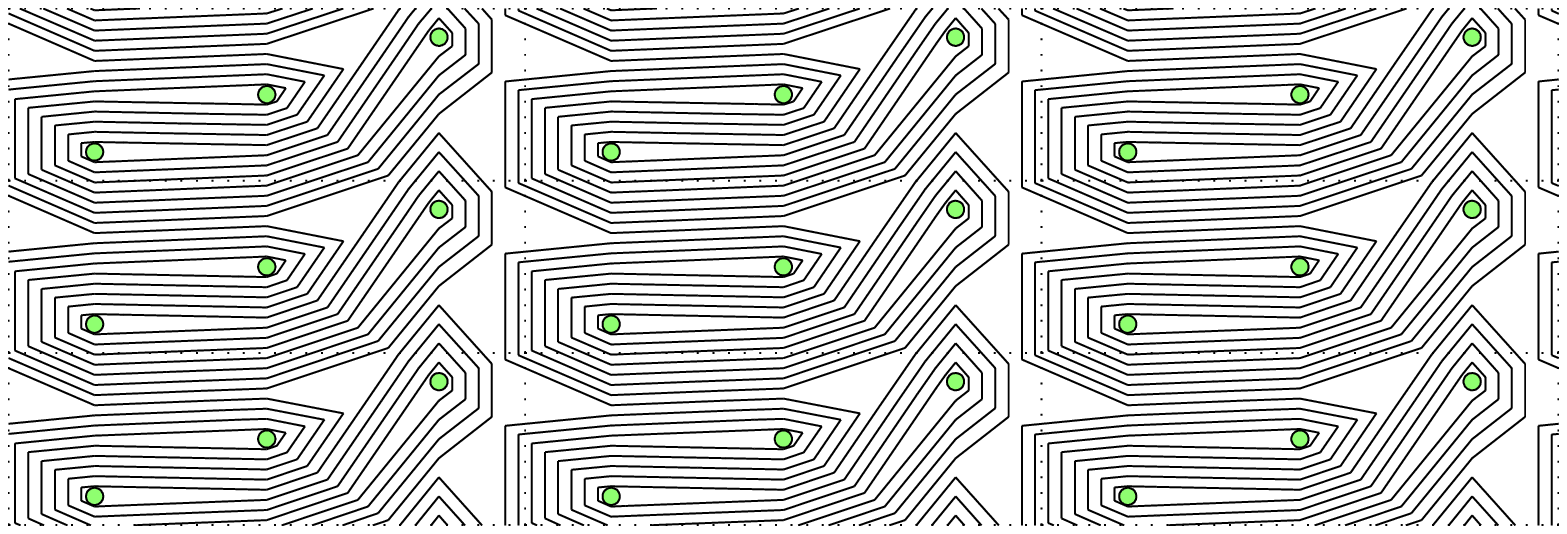} \hspace{0.2cm}
\includegraphics[scale=0.35]{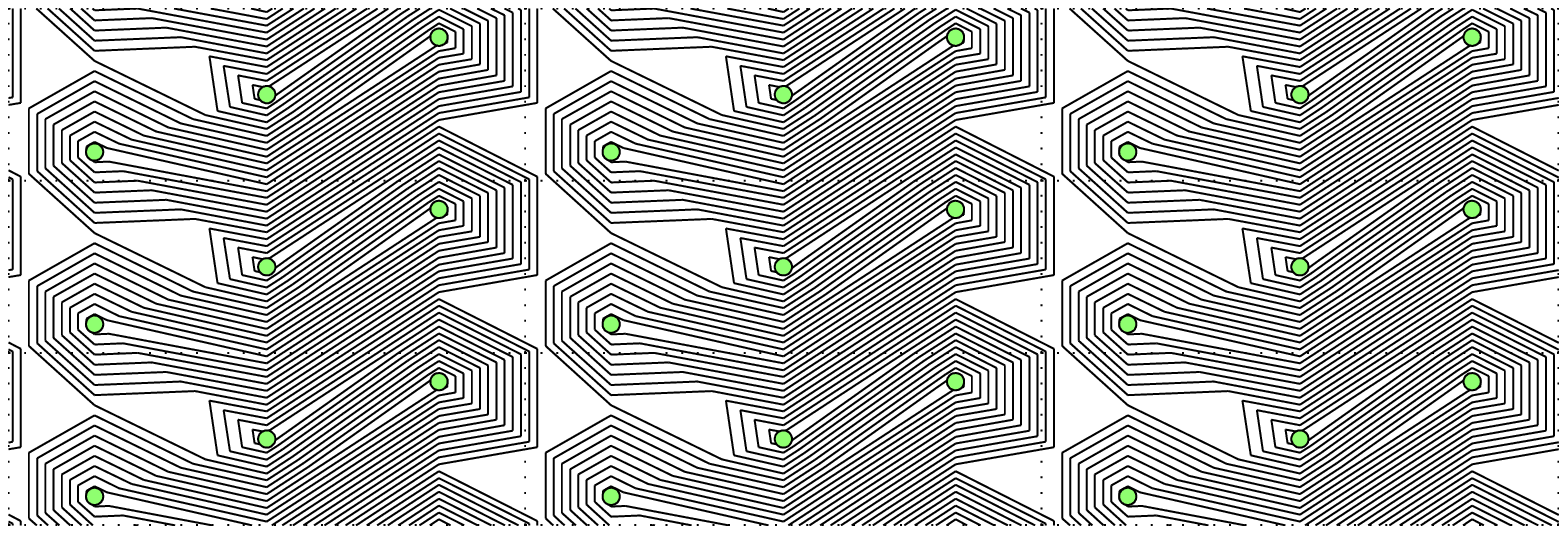} 
\end{center}
\caption{Lamination evolution under the (planar) golden braid 
$\sigma_1 \sigma_2^{-1}$. This braid is proved to have highest topological
entropy per braid letter for a planar braid~\cite{DAlessandro1999}.}
\label{fig:golden}
\end{figure}

The planar pigtail braid $\sigma_1 \sigma_2^{-1}$ with three punctures is
illustrated in Figure~\ref{fig:golden}~\cite{Boyland2000}. This braid is
pseudo-Anosov and has a growth rate per braid letter of $\tsfrac12(1+\sqrt5)$,
which is the golden ratio. This `golden braid' has been proved to have the
highest topological entropy per braid letter for a planar braid
\cite{DAlessandro1999}.

\begin{figure}
\begin{center}
\includegraphics[scale=0.35]{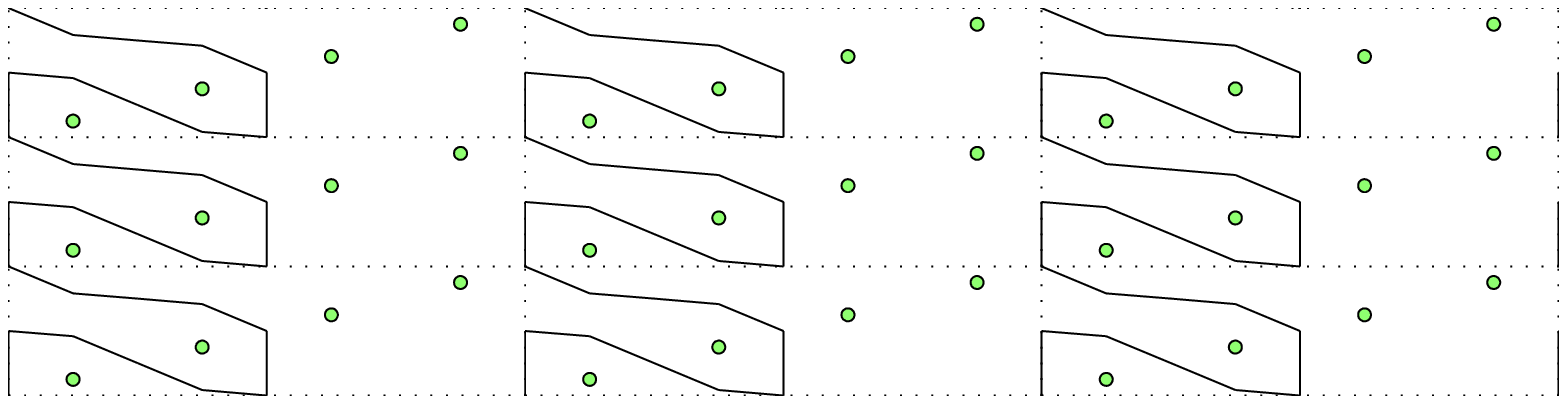} \hspace{0.2cm}
\includegraphics[scale=0.35]{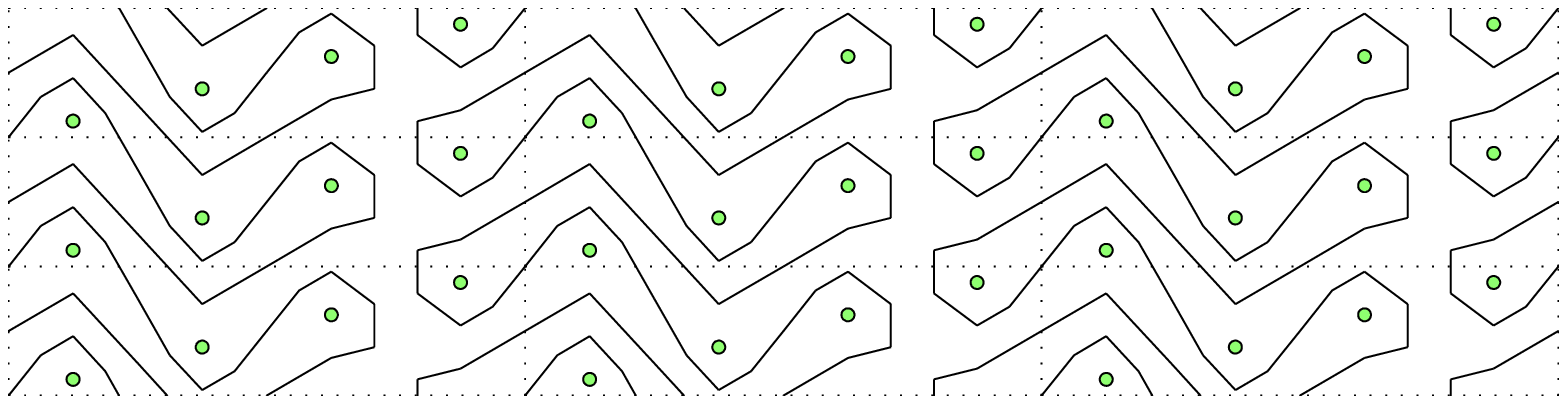}

\vspace{0.2cm}

\includegraphics[scale=0.35]{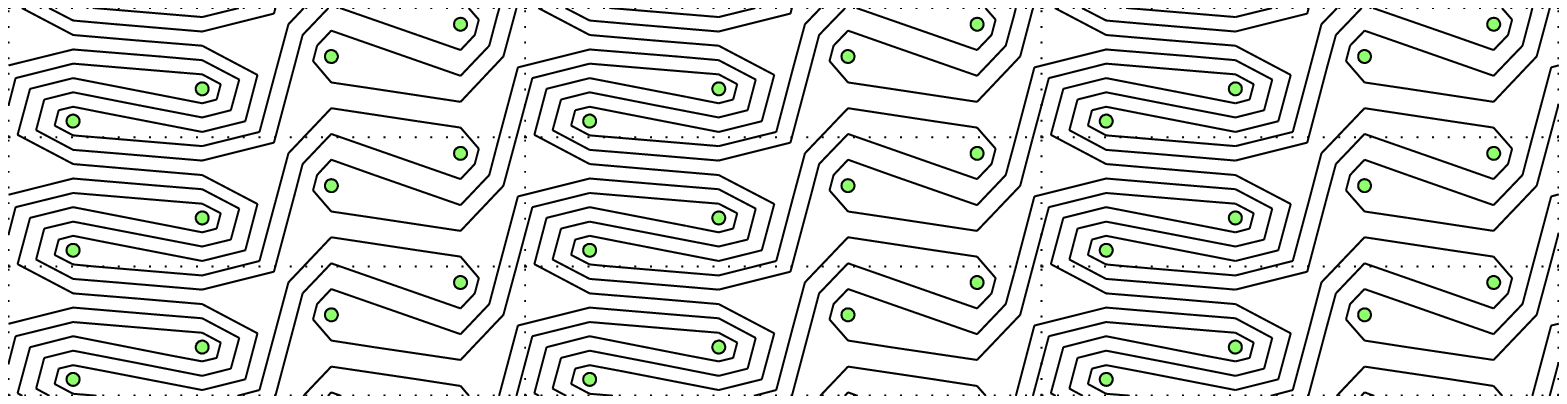} \hspace{0.2cm}
\includegraphics[scale=0.35]{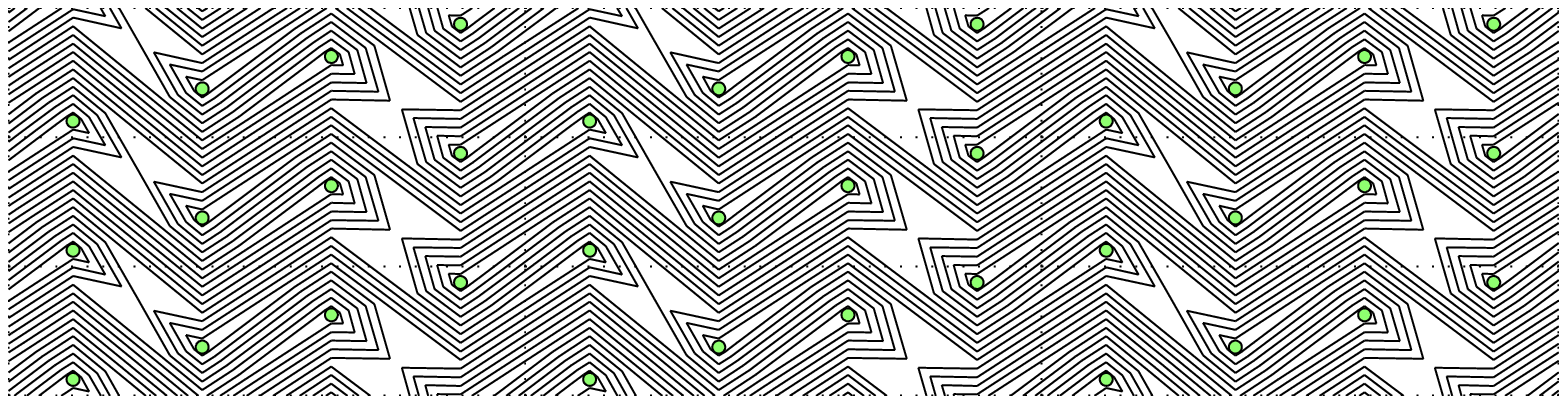}

\vspace{0.2cm}

\includegraphics[scale=0.35]{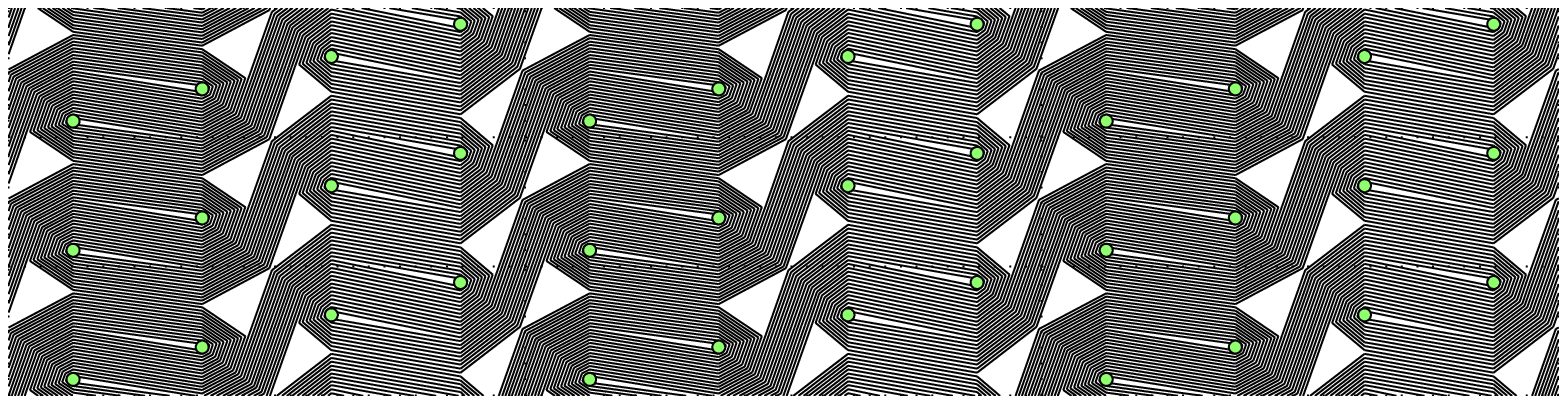} \hspace{0.2cm}
\includegraphics[scale=0.35]{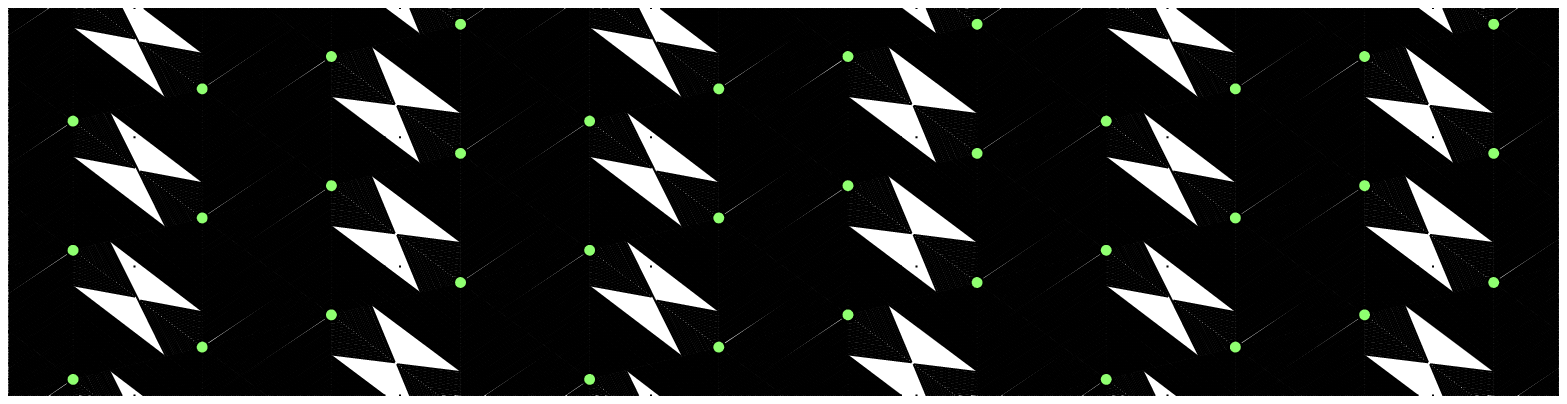}
\end{center}
\caption{Lamination evolution under the cylinder braid $\sigma_1 \sigma_3
\sigma_2^{-1} \sigma_4^{-1}$. Note that the lamination becomes visibly dense
more quickly than for the golden braid in Figure~\ref{fig:golden}.}
\label{fig:silver}
\end{figure}

Figure~\ref{fig:silver} illustrates a related cylinder braid $\sigma_1
\sigma_3 \sigma_2^{-1} \sigma_4^{-1}$ with four punctures.  This is similar to
the pigtail braid but wrapped around a cylinder so that the first and last
punctures are allowed to exchange places.  The cylinder braid $\sigma_1
\sigma_2^{-1}$ with {\it two} punctures has a growth rate per braid letter
given by the silver ratio of $1+\sqrt2$, and this can be proved to be the
optimum per braid letter over all cylinder braids. (The proof follows that of
D'Alessandro~\cite{DAlessandro1999} using a matrix representation for the
cylinder braid group.) The silver braid entropy is almost fifty percent higher
than the golden braid entropy, showing that periodic boundary conditions can
be exploited to enhance chaos. The extra stretching is clearly visible by
comparing Figures~\ref{fig:golden} and \ref{fig:silver}.

\begin{figure}
\begin{center}
\includegraphics[scale=0.35]{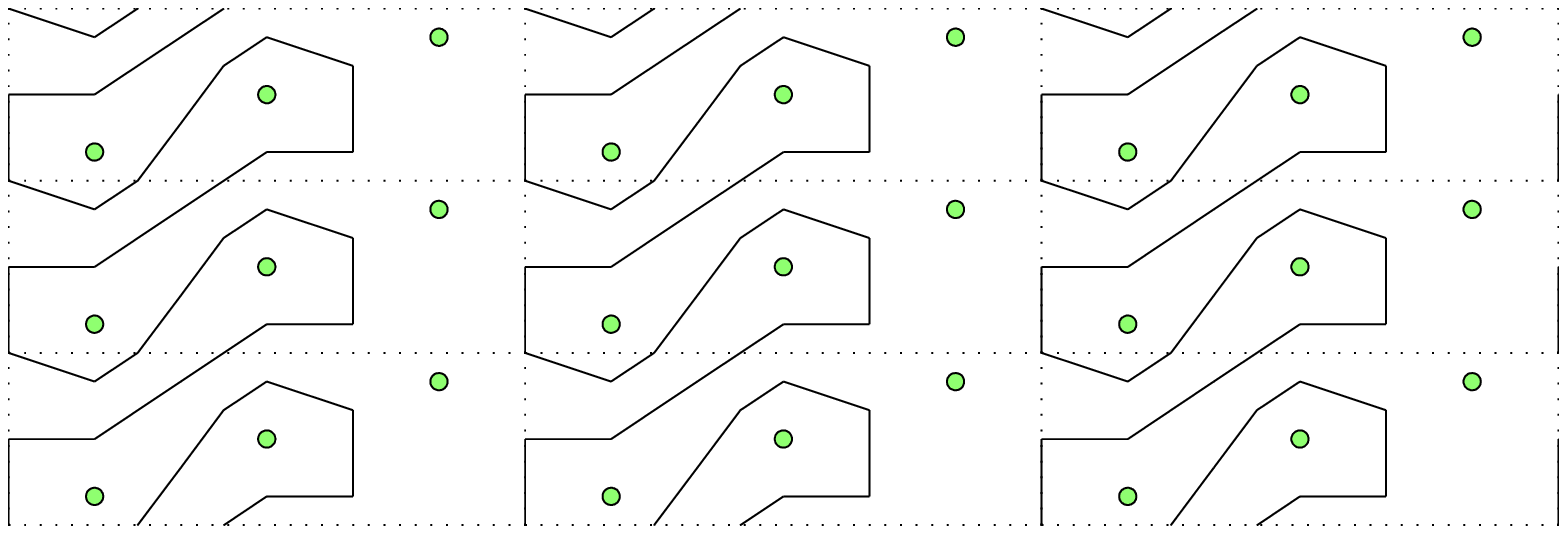} \hspace{0.2cm}
\includegraphics[scale=0.35]{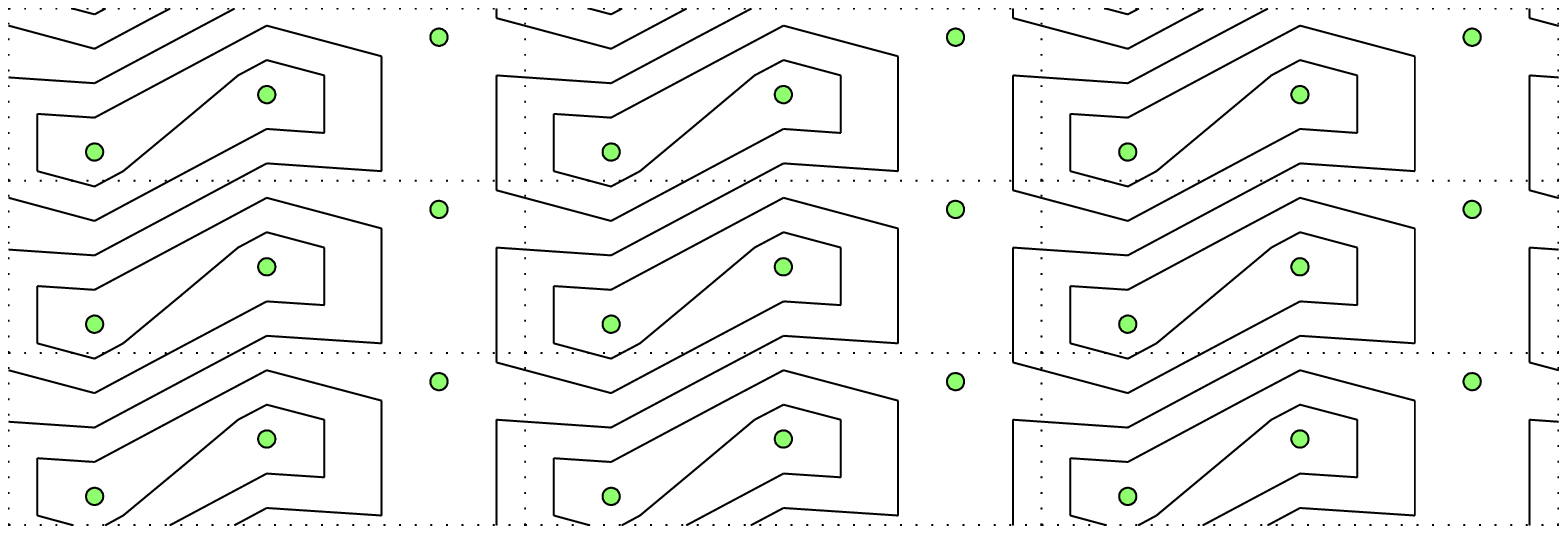}

\vspace{0.2cm}

\includegraphics[scale=0.35]{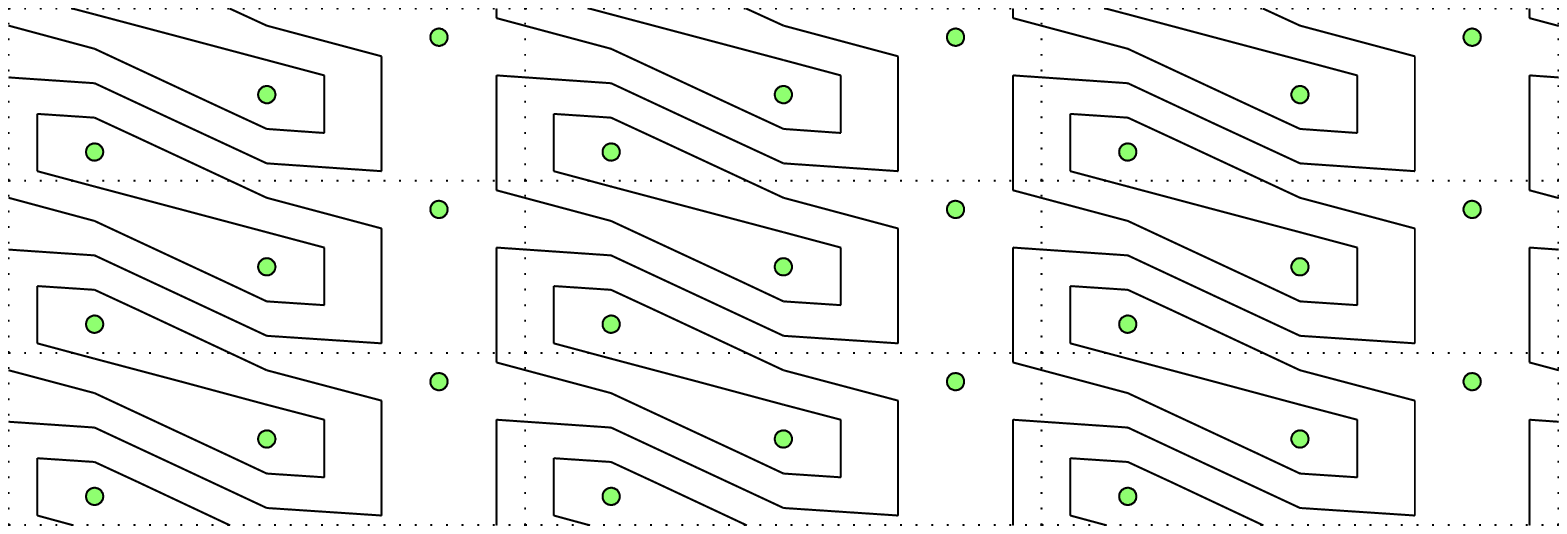} \hspace{0.2cm}
\includegraphics[scale=0.35]{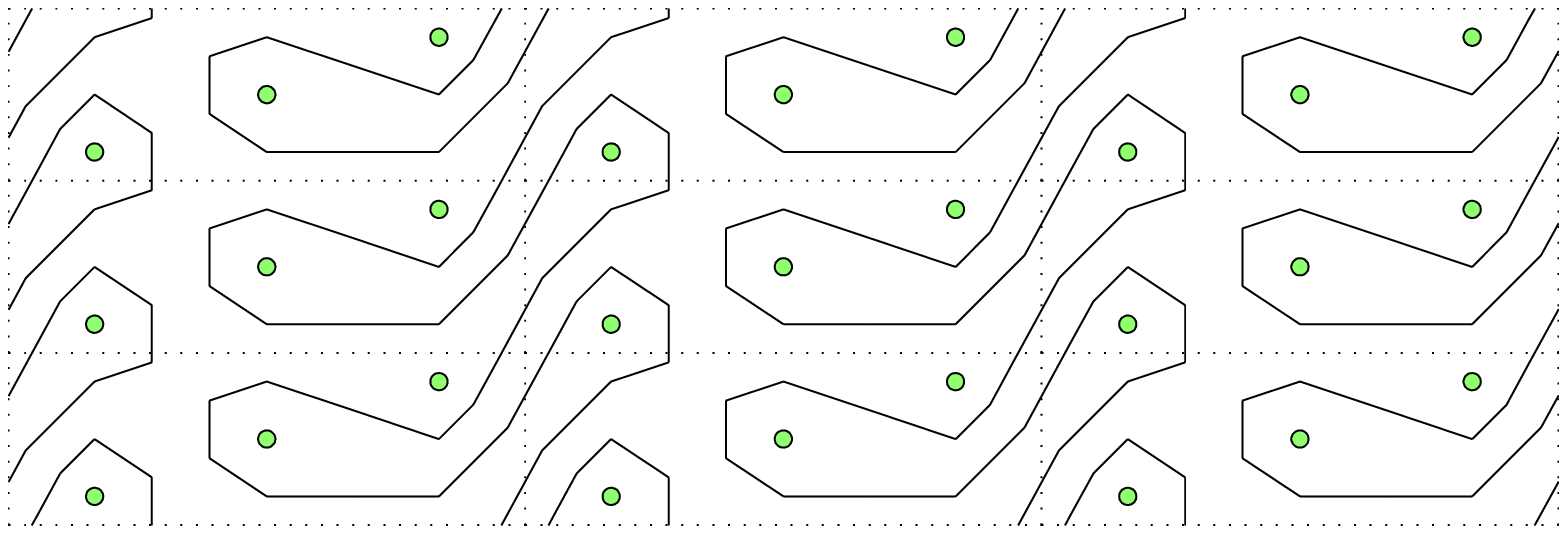}

\vspace{0.2cm}

\includegraphics[scale=0.35]{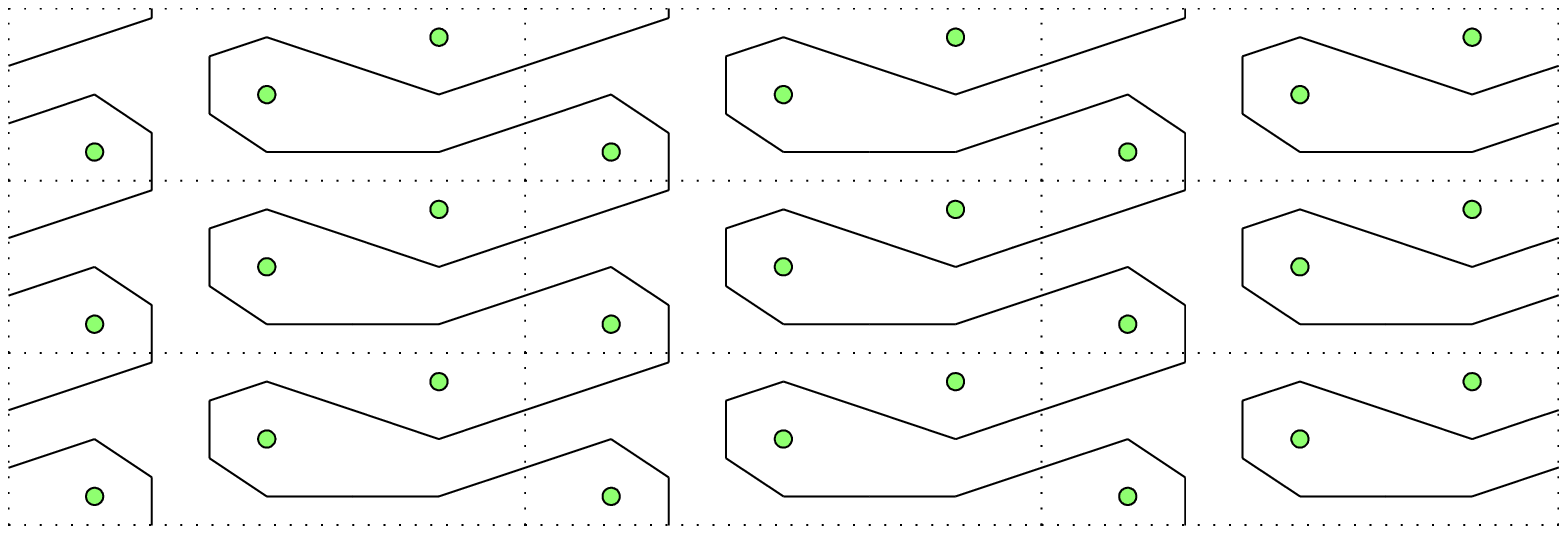} \hspace{0.2cm}
\includegraphics[scale=0.35]{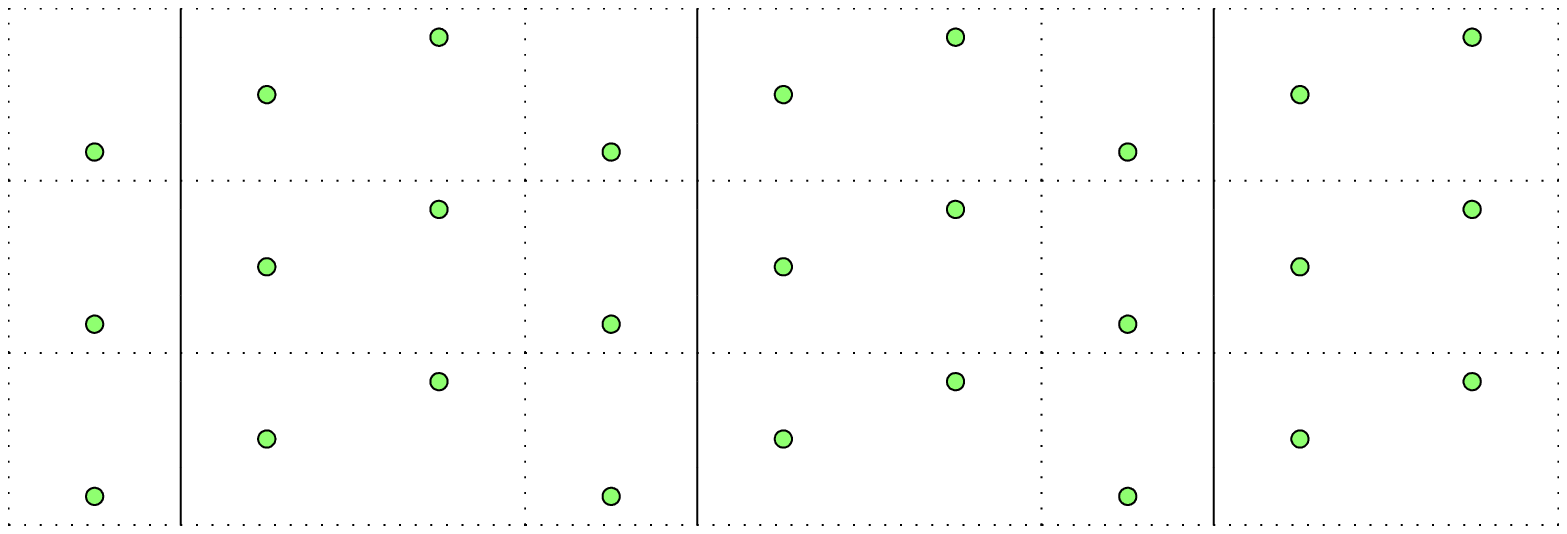}
\end{center}
\caption{Lamination evolution under the toroidal identity braid $\sigma_1^{-2}
\rho_1^{-1} \tau_2 \rho_1 \tau_2^{-1}$. The initial lamination (not shown) is
the vertical line between the first two punctures. The images show, in order,
the lamination after $\sigma_1^{-1}$, $\sigma_1^{-2}$, $\sigma_1^{-2}
\rho_1^{-1}$, $\sigma_1^{-2} \rho_1^{-1} \tau_2 $, $\sigma_1^{-2} \rho_1^{-1}
\tau_2 \rho_1$ and $\sigma_1^{-2} \rho_1^{-1} \tau_2 \rho_1 \tau_2^{-1}$.  The
final lamination is the same as the initial lamination, as required.}
\label{fig:identity}
\end{figure}

A final validation of our encoding and update rules is given by checking that
any lamination is unchanged when it is subjected to the identity braid. For
the braid group on the torus, one way of writing the identity using $\sigma$,
$\rho$ and $\tau$ is $\sigma_1^{-2} \rho_1^{-1} \tau_2 \rho_1 \tau_2^{-1}$
which is taken from the group presentation written down by
Birman~\cite{Birman1969}. In Figure~\ref{fig:identity} the evolution of our
test lamination is shown using this braid, and it is unchanged, as required.

\section{Calculating topological entropies from laminations}
\label{sec:entropy}

Having completed the description of the dynamical system for deformation of a
lamination, we now discuss how the topological entropy of a braid is related
to evolution of a lamination.  The results of Moussafir~\cite{Moussafir2006}
for the punctured sphere apply here: for an appropriate lamination, the growth
rate of the total number of crossings between the lamination and triangulation
converges to the topological entropy the braid~$\lambdastar$.  Thus,
asymptotically we have the estimate
\begin{equation}
\lambdastar = \log  \sum_i \left(x_i^*+y_i^*+z_i^* \right) 
- \log  \sum_i \left( x_i+y_i+z_i \right),
\label{eq:estimate}
\end{equation}
where the starred variables are the updated crossing numbers described in
Section~\ref{sec:deformation}. Convergence of $\lambdastar$ to the exact braid
entropy $\lambda$ appears to be exponential (provided the braid has a
pseudo-Anosov component), and so in practice very few iterations are required
to obtain $\lambda$ to double precision. Finite-order braids, which have zero
topological entropy are detected easily by checking for sub-exponential
convergence of $\lambdastar$.

As an example, the exact topological entropy of the four-strand braid 
$\sigma_1 \sigma_3 \sigma_2^{-1} \sigma_4^{-1}$ (see the previous section and
Figure~\ref{fig:silver}) 
is known to be twice the 
logarithm of the silver ratio $\lambda = 2 \log (1+\sqrt{2}) = 
1.762747174039086\dots$. In Table~\ref{tab:growth} we show how the total 
crossing number and the entropy estimate $\lambdastar$ given by 
\eqref{eq:estimate} evolve under the first few iterations of the braid 
(using the initial lamination described in Section~\ref{sec:examples}). 
Double precision accuracy of the entropy, which is sufficient for most 
purposes, is reached after about 20 iterations.

\begin{table}
\begin{center}
\begin{tabular}{rrrrrrr} \hline
iteration & total crossings & entropy $\lambdastar$ & error
$|\lambda-\lambdastar|$ \\ \hline
1  & 24                   &    2.48490664978800     &  0.72215947574891    \\
2  & 154                  &    1.85889877206568     &  0.09615159802660    \\
3  & 912                  &    1.77868738766070     &  0.01594021362162    \\
4  & 5330                 &    1.76546652708556     &  0.00271935304647    \\
5  & 31080                &    1.76321328732169     &  0.00046611328261    \\
6  & 181162               &    1.76282713309230     &  0.00007995905321    \\
7  & 1055904              &    1.76276089245107     &  0.00001371841199    \\
8  & 6154274              &    1.76274952773491     &  0.00000235369583    \\
9  & 35869752             &    1.76274757786911     &  0.00000040383002    \\
10 & 209064250            &    1.76274724332535     &  0.00000006928627    \\
11 & 1218515760           &    1.76274718592673     &  0.00000001188765    \\
12 & 7102030322           &    1.76274717607868     &  0.00000000203960    \\
13 & 41393666184          &    1.76274717438903     &  0.00000000034994    \\
14 & 241259966794         &    1.76274717409913     &  0.00000000006004    \\
15 & 1406166134592        &    1.76274717404939     &  0.00000000001030    \\
16 & 8195736840770        &    1.76274717404085     &  0.00000000000177    \\
17 & 47768254910040       &    1.76274717403939     &  0.00000000000030    \\
18 & 278413792619482      &    1.76274717403914     &  0.00000000000005    \\
19 & 1622714500806864     &    1.76274717403910     &  0.00000000000001    \\
20 & 9457873212221714     &    1.76274717403908     &  0.00000000000000    \\
\hline
\end{tabular}                                           
\end{center}
\caption{Convergence of the entropy estimate $\lambdastar$ in
\eqref{eq:estimate} towards the exact topological entropy $\lambda = 2 \log
(1+\sqrt{2}) = 1.762747174039086\dots$ for the silver braid
$\sigma_1 \sigma_3 \sigma_2^{-1} \sigma_4^{-1}$.  Convergence appears to be
exponential for a pseudo-Anosov braid, with approximately one extra digit per
iteration~\cite{Moussafir2006}.}
\label{tab:growth}
\end{table}

Braids quite naturally arise when considering periodic trajectories of points
in a two-dimensional flow. For a spatially-periodic flow the space-time plot
of these trajectories is typically a cylinder or a torus braid, and the
entropy of this braid provides a rigorous lower bound on the topological
entropy of the flow~\cite{Boyland1994,Boyland2000}.  Finn et
al.~\cite{Finn2005} derived such a lower bound for the sine flow using
Thurston's `iterate and guess' method to construct the train-track for the
braid formed by a set of periodic orbits. We will now describe how the same
result can be found using our lamination approach.

The sine flow is a time-periodic alternating shear-flow defined on the unit
torus \hbox{$0\le x,y<1$}. The period is~$\sineparam$, with the velocity field
given by $(\sin 2\pi y,0)$ for {$0\le t<\sineparam/2$} and $(0,\sin 2\pi x)$
for \hbox{$\sineparam/2 \le t < \sineparam$}, with~$t$ marking time. This
simple flow has been well studied because the parameter range $0 \le
\sineparam \lesssim 2$ gives rich dynamics that vary from complete
integrability to almost global chaos (with few visible islands in a Poincar\'e
section).

Since the flow is piecewise steady it is easy to construct a map to track the
motion of points from one period to the next. Consequently, the entropy of 
the flow for a given $\sineparam$ can be found quickly and simply by direct
numerical simulation of line stretching.  Once the entropy of the flow is 
known, it is instructive to see what prediction of the entropy is given by 
considering the braiding of a finite number of particle orbits. 

\begin{figure}
\begin{center}
\includegraphics[width=.6\textwidth]{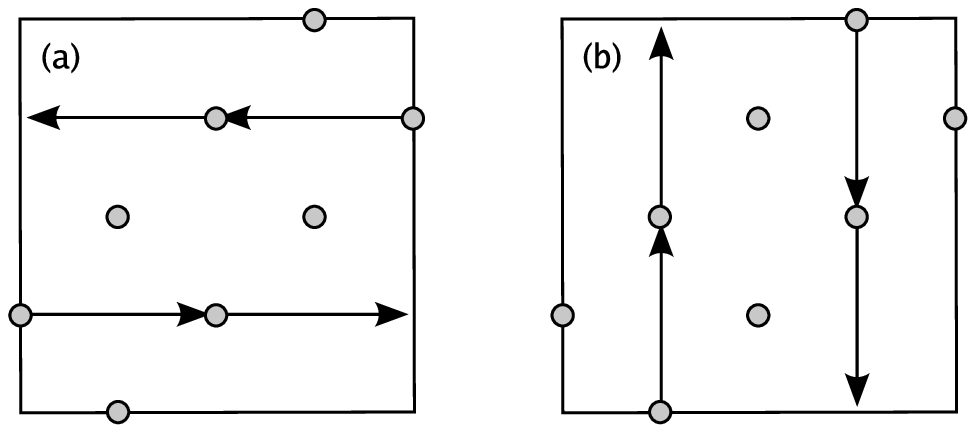}
\end{center}
\caption{The four periodic orbits considered for the sine flow
  with~$\sineparam=1$.  (a) $\{(0,\tsfrac14),(\tsfrac12,\tsfrac14)\}$ and
  $\{(\tsfrac12,\tsfrac34),(1,\tsfrac34)\}$; (b)
  $\{(\tsfrac14,0),(\tsfrac14,\tsfrac12)\}$ and
  $\{(\tsfrac34,\tsfrac12),(\tsfrac34,1)\}$.}
\label{fig:sineflow_orbits}
\end{figure}

For general $\sineparam$ it is difficult to locate unstable periodic orbits
due to the highly chaotic nature of the sine mapping; however, for the special
parameter $\sineparam=1$ it is quite easy to spot some of the low order
orbits.  In particular, four period-two orbits are given by
$\{(0,\tsfrac14),(\tsfrac12,\tsfrac14)\}$,
$\{(\tsfrac12,\tsfrac34),(1,\tsfrac34)\}$,
$\{(\tsfrac14,0),(\tsfrac14,\tsfrac12)\}$, and
$\{(\tsfrac34,\tsfrac12),(\tsfrac34,1)\}$.  The first pair of orbits is
depicted in Figure~\ref{fig:sineflow_orbits}(a), the second pair in
Figure~\ref{fig:sineflow_orbits}(b).  The points associated with the first
pair of periodic orbits do not move in the second half-period, while those
associated with the second pair do not move in the first half-period.
\begin{figure}
\begin{center}
\includegraphics[width=\textwidth]{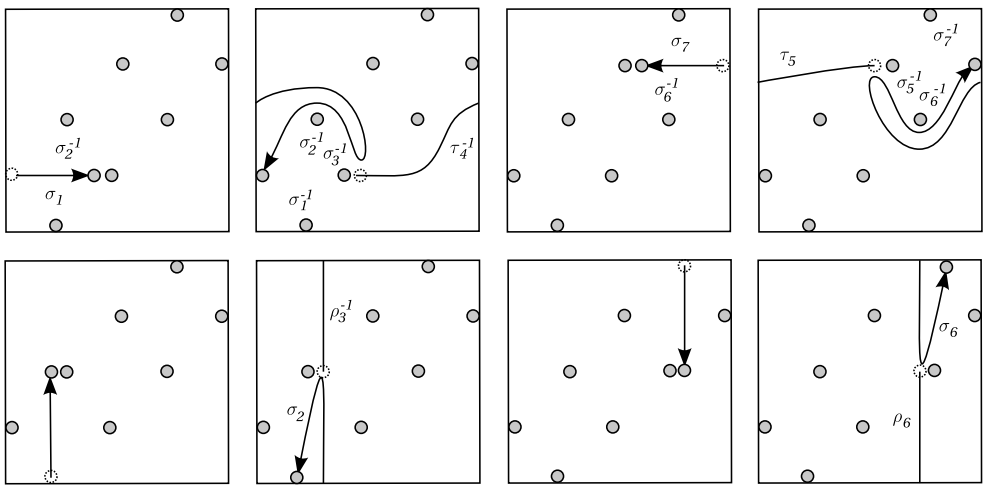}
\end{center}
\caption{Illustration of the eight strand braid formed by a set of period two
  points in the sine flow with $\sineparam=1$. The braid word is $\sigma_1
  \sigma_2^{-1} \tau_4^{-1} \sigma_3^{-1} \sigma_2^{-1} \sigma_1^{-1} \sigma_7
  \sigma_6^{-1} \tau_5 \sigma_5^{-1} \sigma_6^{-1} \sigma_7^{-1} \rho_3^{-1}
  \sigma_2 \rho_6 \sigma_6$. This braid has been shown to have an exact
  entropy of $1.21875572687\dots$ using train-tracks~\cite{Finn2005}. The
  entropy estimate using the growth of laminations agrees to every decimal
  place calculated.}
\label{fig:sineflow}
\end{figure}

Now we must determine the braid formed by the eight trajectories forming these
four periodic orbits.  We first disambiguate the order of the periodic points
by displacing them slightly along the~$x$ axis, as shown in
Figure~\ref{fig:sineflow}.  Then we encode the trajectories in terms of braid
group generators, deforming as needed.  Deforming is necessary since usually
the trajectory does not map directly onto a generator, and some intermediate
operations must be insterted.  For instance, in the second snapshot in
Figure~\ref{fig:sineflow} the generators $\sigma_3^{-1} \sigma_2^{-1}
\sigma_1^{-1}$ are used to return the point to the leftmost position after a
$\tau_4^{-1}$ operation.  After a full period, the resulting braid word is
$\sigma_1 \sigma_2^{-1} \tau_4^{-1} \sigma_3^{-1} \sigma_2^{-1} \sigma_1^{-1}
\sigma_7 \sigma_6^{-1} \tau_5 \sigma_5^{-1} \sigma_6^{-1} \sigma_7^{-1}
\rho_3^{-1} \sigma_2 \rho_6 \sigma_6$.  Using equation (\ref{eq:estimate}) we
find the entropy of the braid converges to $1.21875572687\dots$, which agrees
with an alternative calculation using train-tracks~\cite{Finn2005}. The
entropy lower bound given by the braid accounts for $82\%$ of the flow entropy
of approximately $1.48$.

Since it is difficult to find any exact periodic orbits for general
$\sineparam$ it is natural to ask whether the entropy can be found by
considering the braiding of any selection of trajectories. Since point motions
in the sine flow are piecewise horizontal and vertical, and, for the purposes
of braiding, can be performed sequentially, it is very easy and
computationally fast to calculate all the $\sigma$, $\rho$ and $\tau$
operations that occur during each half-period of the flow, for an arbitrary
number of points.

For the first half-period, where all motions are horizontal, we record a
$\sigma$ operation for each change in order of the particle $x$ coordinates.
The sign of each crossing is determined by the difference in $y$ coordinates
for the two points that cross. Special attention is required when a point
crosses the periodic boundary. Crossing leftwards over $x=0$ is achieved by
$\tau_1 \sigma_1^{-1} \cdots \sigma_{n-1}^{-1}$; this moves the leftmost
point one copy of the domain to the left, and then, through a
sequence of $\sigma$ operations, moves the point to position $n$ and undoes
all the undesired crossings with the other points. Similarly, when the
rightmost point crosses over $x=1$ the string $\tau_n^{-1} \sigma_{n-1}^{-1}
\cdots \sigma_{1}^{-1}$ is assigned.  The computation is simpler during the
second half-period: all point motions are vertical, so no $\sigma$ motions
occur. If the $i$th point in the $x$ direction crosses downwards through $y=0$
this is labelled $\rho_i$. Likewise, an upwards crossing of $y=1$ is assigned
$\rho_i^{-1}$.

\begin{figure}
\begin{center}
\includegraphics[width=10cm]{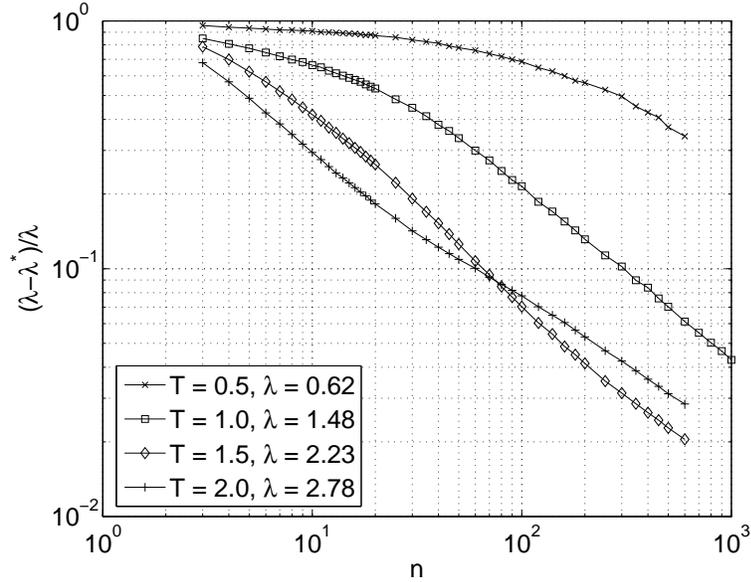}
\end{center}
\caption{Convergence of the mean entropy estimate $\langle \lambdastar
\rangle$ towards the exact entropy $\lambda$ as the number of trajectories $n$
in the braid is increased. For large $n$ the error appears to decrease
according to a power law. For smaller $\sineparam$ the convergence is hindered
by many of the points being inside periodic islands.}
\label{fig:sineentropy}
\end{figure}

To estimate the entropy of the flow we evolve a lamination according to 
the braid that results from the point motions. Since the points are not 
periodic, the growth per iteration continues to vary, but the average 
value of the growth always converges. The initial points are chosen at 
random throughout the domain, and can either live in a chaotic region or 
in a periodic island. Hence the predicted entropy for a given number of 
points may depend on the initial positions. To allow for this we use an 
estimate $\langle \lambdastar \rangle$ obtained by averaging over many 
realizations with different initial points.

In Figure~\ref{fig:sineentropy} we show how the braid entropy $\langle
\lambdastar \rangle$ converges towards the flow entropy $\lambda$ as the
number of trajectories in the braid is increased. For the four different
values of $\sineparam$ we have considered, convergence appears to be as a
power law for large $n$.  Although we are unable to prove that the exact
entropy is reached in the limit as $n \rightarrow \infty$ we would expect this
since in this limit the braid will contain perfect information about the flow.
For small values of $\sineparam$ (such as $\sineparam=0.5$ in
Figure~\ref{fig:sineentropy}) the flow contains large islands of regularity
which do not contribute greatly to line stretching. In this regime~$n$ has to
become relatively large before there are enough points exploring and encoding
the dynamics in the small chaotic region.

Owing to the efficiency with which the braid is determined and analyzed, we
are able to consider much larger times and number of points $n$ than has been
considered in previous articles~\cite{Thiffeault2005}.  Another point worth
noting is that for $\sineparam=1$, the mean entropy estimate $\langle
\lambdastar \rangle = 0.45$ with $n=8$ eight points is much worse than the
estimate of $\lambdastar = 1.22$ found by using the set of eight judiciously
chosen periodic points considered earlier. This highlights the important role
of low order periodic orbits in determining much of the nature of the flow.

\section{Discussion}
\label{sec:discussion}

We have derived a dynamical system to compute the evolution of a lamination
(equivalence class of simple closed curves) under the braiding of an arbitrary
number of punctures on the torus. The method is essentially a modification of
the Dynnikov coordinate approach employed by Moussafir~\cite{Moussafir2006},
but we use a triangulation encoding that has favorable properties for studying
torus braids.  Naturally, our method also works for the special cases of
cylinder and planar braids.  However, in the planar case our dynamical system
still has~$3n$ variables, more than the~\hbox{$2n-3$} required by Moussafir,
so there is clearly some redundancy in this case.  Also, the triangulation of
our domain is not unique, so we expect that the details of our method are not
unique.  However our triangulation seems to be the best choice for simplifying
the arithmetic.

For completeness we point out that our dynamical system does not work directly
for~\hbox{$n=2$} punctures because in equation (\ref{eq:sigmaup}) and
(\ref{eq:sigmaiup}) this would mean that the indices~\hbox{$i-1$}
and~\hbox{$i+1$} refer to the same quantities. There is nothing difficult
about the case $n=2$ though, and in principle one could write down the
corrected update rules
for~\hbox{$\{x_1,y_1,z_1,x_2,y_2,z_2\}$}. Alternatively, a lazy but convenient
workaround is to include a redundant third puncture, glued to one of the other
two, so that it is slaved to its motion.  At a little more computational
expense this allows one computer code to handle all values of~\hbox{$n\ge2$}.
The case $n=1$ is trivial.

By using similar arguments to Moussafir~\cite{Moussafir2006}, it can be shown
that as the number of iterations tends to infinity the logarithm of the number
of lamination crossings grows at the rate of the braid topological
entropy. Though the number of crossings grows exponentially fast, our
numerical implementation of the dynamical system uses a large integer
arithmetic library to allow calculation of the entropy to arbitrary
precision. Convergence appears to be exponentially fast, with approximately
one digit of accuracy gained per iteration for the braids we have considered.

In practice, if only a few digits of accuracy are required then double
precision floating-point arithmetic for the crossing numbers is adequate and
can speed up code significantly.  A caveat with using floating-point
arithmetic is that is destroys reversibility.  In general by performing a long
braid followed by its inverse the initial lamination will not be recovered due
to exponential growth of small roundoff errors.  This is akin to
irreversibility due to numerical diffusion in trajectory computations in a
chaotic Stokes flow.  However, since the dynamical system itself is exact, the
only errors are due to roundoff and not to discretization, so even with double
precision the dynamical system is surprisingly reversible.

If finite precision is not acceptable, our method is easily adapted to find
exact entropies.  This can be done by shortcircuiting the minimum functions in
the update rules (\ref{eq:rhoup})--(\ref{eq:sigmaiup}).  Under repeated
application of a braid, the dynamical system quickly converges to a periodic
pattern where it is known in advance which number will be taken for each
minimum. With this knowledge each call to the min function can be replaced by
the correct variable and then a linear system can be written down for the
crossing numbers. The logarithm of the modulus of the largest eigenvalue will
give the exact entropy.

This work was motivated by the study of two-dimensional fluid mixing via the
braiding motion of fluid particle
trajectories~\cite{Gambaudo1999,Thiffeault2005}. In this setting we have
derived an efficient tool that allows practical analysis of large braids.
Outside of this particular application it is natural to ask whether the method
can be generalized for the braid group on surfaces of higher genus, as
considered by Birman~\cite{Birman1969}.  Since any surface can be
triangulated, in principle our method could be extended to higher genus.  The
problem anticipated with generalizing the approach is that it is more
difficult to visualize the preimage problems on a surface with many holes, and
also it is not clear in general how to exploit group properties to gain a
complete set of crossing update rules.

We thank Colin Cotter and Jacques-Oliver Moussafir for helpful
discussions. This work was funded by the UK Engineering and Physical Sciences
Research Council grant GR/S72931/01.


\end{document}